\documentclass[11pt]{article}
\usepackage{amsmath,amsfonts,amsthm}
\newtheorem{theorem}{Theorem}[subsection]
\newtheorem{corollary}[theorem]{Corollary}
\newtheorem{lemma}[theorem]{Lemma}

\newtheorem{proposition}[theorem]{Proposition}
\newtheorem{definition}[theorem]{Definition}

\newtheorem{fact}[theorem]{Fact}

\def\orig#1{\bar{#1}}
\def\calN{\mathcal{N}}

\def\form#1#2{#1^{T}#2}

\newcommand{\ceiling}[1]{\left\lceil#1\right\rceil}

\newcommand\symProb{\operatorname{\textbf{Pr}}\displaylimits}
\newcommand\symExpec{\operatorname{\textbf{E}}\displaylimits}
\def\prob#1#2{\symProb_{#1}\left[ #2 \right]}
\def\expec#1#2{\symExpec_{#1}\left[ #2 \right]}

\def\norm#1{\left\| #1 \right\|}
\def\fnorm#1{\left\| #1 \right\|_{F}}
\def\setof#1{\left\{#1  \right\}}

\def\dist#1#2{\mbox{{\bf dist}}\left(#1, #2 \right)}

\def\bdry#1{\mbox{\bf bdry} (#1)}

\def\pos#1{\mathcal{H} (#1)}

\def\aa{\pmb{\mathit{a}}}
\newcommand\bb{\boldsymbol{\mathit{b}}}
\newcommand\cc{\boldsymbol{\mathit{c}}}
\newcommand\CC{\boldsymbol{\mathit{C}}}
\newcommand\ee{\boldsymbol{\mathit{e}}}

\newcommand\KK{\boldsymbol{\mathit{K}}}
\newcommand\pp{\boldsymbol{\mathit{p}}}
\newcommand\qq{\boldsymbol{\mathit{q}}}
\newcommand\rr{\boldsymbol{\mathit{r}}}
\renewcommand\SS{\boldsymbol{\mathit{S}}}

\newcommand\xx{\boldsymbol{\mathit{x}}}
\newcommand\yy{\boldsymbol{\mathit{y}}}
\newcommand\zz{\boldsymbol{\mathit{z}}}
\def\tt{\boldsymbol{\mathit{t}}}

\def\bvec#1{{\mbox{\boldmath $#1$}}}

\def\cone#1{\mbox{{\bf Cone}}\left(#1  \right)}
\def\ray#1{\mbox{{\bf Ray}}\left(#1  \right)}
\def\vp#1#2#3{#1_{#2},\ldots , #1_{#3}}
\def\vs#1#2#3{#1_{#2},\ldots , #1_{#3}}

\def\hull#1{\mbox{{\bf Hull}}\left(#1  \right)}

\def\abs#1{\left|#1  \right|}
\def\intersect{\cap}
\def\intersects{\mbox{{ intersects }}}

\newdimen\pIR
\newcommand\StevesR{{\rm I\kern\pIR R}}
\def\Reals#1{\StevesR^{#1}}

\begin{document}

\title{Smoothed Analysis of Renegar's Condition Number for Linear Programming}

\author{
John Dunagan\thanks{Supported in part by NSF Grant CCR-9875024.
{\texttt{jdunagan@microsoft.com}}. Part of the work was done while the
author was at MIT.}\\
Microsoft Research
\and
Daniel A. Spielman
\thanks{Partially supported by an Alfred P. Sloan Foundation Fellowship,
and NSF Grant CCR-0112487.
\texttt{spielman@math.mit.edu}}\\
Department of Mathematics\\
Massachusetts Institute of Technology
\and
Shang-Hua Teng
\thanks{
Partially supported by an Alfred P. Sloan Foundation Fellowship,
and NSF Grants CCR-9972532, and CCR-0112487. \texttt{steng@cs.bu.edu}}\\
Department of Computer Science\\
Boston University and\\
Akamai Technologies Inc.}

\maketitle

\begin{abstract}
We perform a smoothed analysis of Renegar's condition number
  for linear programming.
In particular, we show that for every $n$-by-$d$ matrix
  $\orig{A}$, $n$-vector $\orig{\bb}$ and $d$-vector $\orig{\cc}$
  satisfying $\fnorm{\orig{A}, \orig{\bb}, \orig{\cc}} \leq 1$
  and every $\sigma \leq 1/\sqrt{dn}$,
  the expectation of the logarithm of $C (A,\bb ,\cc)$ is
  $O (\log (n d / \sigma ))$, where
  $A$, $\bb$ and $\cc$ are Gaussian perturbations of
  $\orig{A}$, $\orig{\bb}$ and $\orig{\cc}$ of variance $\sigma^{2}$.
From this bound, we obtain a smoothed analysis of Renegar's
  interior point algorithm.
By combining this with the smoothed analysis of finite
   termination Spielman and Teng (Math. Prog. Ser. B, 2003),
  we show that the smoothed complexity of linear programming
  is $O (n^{3} \log (n d / \sigma ))$.
\end{abstract}

\section{Introduction}\label{sec:intro}

In~\cite{SpielmanTeng}, Spielman and Teng introduced the
  smoothed analysis of algorithms as an alternative to
  worst-case and average-case analyses
  in the hope that it could provide a measure of the
  complexity of algorithms that better agrees
  with practical experience.
The smoothed complexity of an algorithm is the maximum over
  its inputs of the expected running time of the
  algorithm under slight perturbations of that input.
In this paper,
  we perform a smoothed analysis of Renegar's condition
  number for linear programs, and thereby obtain
  a smoothed analysis of his interior-point algorithm.
Interior point algorithms for linear programming are
  exciting both because they are known to run
  in polynomial time~\cite{Karmarkar} in the worst case
  and because they have been used to efficiently
  solve linear programs in practice.
In fact, the speed of interior point methods in practice
  is much better than that proved in their worst-case
  analyses~\cite{LustigMarstenShanno3,LustigMarstenShanno2,AndersenIPM}.
This discrepancy between worst-case analysis and practical experience
  is our main motivation for studying the smoothed complexity
  of interior point methods.

Our main result is that the smoothed value of
  Renegar's condition number, to be defined in Section~\ref{ssec:RenegarsWork},  
  is $O (\log (n d / \sigma ))$.
That is, for each $(\orig{A},\orig{\bb},\orig{\cc })$ and
  $\sigma \leq 1/\sqrt{dn}$, 
\[
\expec{(A,\bb ,\cc )
      \leftarrow \calN  ((\orig{A},\orig{\bb},\orig{\cc }),\sigma )}
     {C (A, \bb , \cc ) } = O (\log (n d / \sigma )),
\]
where $\calN ((\orig{A},\orig{\bb},\orig{\cc }),\sigma )$
  is the distribution of Gaussian perturbations
  of $(\orig{A},\orig{\bb},\orig{\cc })$ of variance $\sigma^{2}$, and
  $(A,\bb ,\cc )
      \leftarrow \calN  ((\orig{A},\orig{\bb},\orig{\cc }),\sigma )$
  indicates that $(A, \bb, \cc )$ is chosen according to this distribution.
As Renegar's algorithm~\cite{RenegarFunc} takes 
  $O \left(\sqrt{n} \ln \left(C (A, \bb ,\cc )/\epsilon  \right)\right)$
  iterations to find a solution of relative
  accuracy $\epsilon $,
  we find that the smoothed complexity of Renegar's algorithm 
  when it is asked for a solution of relative accuracy $\epsilon $
  is $O (n^{3} \log (n d / \sigma \epsilon ))$.

As explained in~\cite{SpielmanTengTermination}, when one combines
  this analysis with the smoothed analysis of the finite termination
  procedure in that paper, one obtains an interior point
  algorithm that returns the exact answer to the linear program
  and has smoothed complexity 
  $O (n^{3} \log (n d / \sigma ))$.
In comparison, the best-known bound on the
   worst-case complexity of any linear programming algorithm
   is Vaidya's~\cite{Vaidya} bound of 
   $O ((n+d)d^{2} + (n+d)^{1.5}d)L)$,
  and the best known bound for an interior point method is
  $O (n^{3} L)$, first due to Gonzaga~\cite{Gonzaga}.

\subsection{The Complexity of Linear Programming Algorithms}

A linear program is typically specified by a matrix $A$ together with
  two vectors $\bb $ and $\cc $.
If $A$ is an $n$ by $d$ matrix, then $\bb$ is an $n$-vector and
  $\cc $ is a $d$-vector.
There are several canonical forms of linear programs
  specified by $(A,\bb,\cc)$.
The following are four commonly used canonical forms:

\begin{eqnarray}
  \max~ \cc^{T}\xx  \mbox{ s.t. } A\xx \leq \bb
    & \mbox{ and its dual } &
  \min~ \bb^{T}\yy~\mbox{ s.t } A^{T}\yy = \cc,~~ \yy \geq \bvec{0}\label{eqn:form1}
%\makebox[0in]{\hspace{0.375in} (1)}
\\
 \max~ \cc^{T}\xx ~ \mbox{ s.t. } A \xx  \leq \bb, ~ \xx  \geq \bvec{0}
  & \mbox{ and its dual } &
\min~ \bb^{T} \yy ~ \mbox{ s.t. } A^{T} \yy  \geq \cc , ~ \yy  \geq
\bvec{0} \label{eqn:form2}
%\makebox[0in]{\hspace{.77in} (2)}
\\
 \max~ \cc^{T}\xx ~ \mbox{ s.t. } A \xx  = \bb, ~ \xx  \geq \bvec{0}
  & \mbox{ and its dual } &
\min~ \bb^{T} \yy ~ \mbox{ s.t. } A^{T} \yy  \geq \cc \label{eqn:form3}
%\makebox[0in]{\hspace{1.78in} (3)}
\\
 \mbox{ find $\xx \not = \bvec{0} $ s.t. } A \xx  \leq \bvec{0}
  & \mbox{ and its dual } &
 \mbox{find $\yy \not = \bvec{0}$ s.t. }
         A^{T} \yy = \bvec{0} , ~ \yy  \geq \bvec{0} \label{eqn:form4}
%\makebox[0in]{\hspace{.59in} (4)}
\end{eqnarray}

Without loss of generality, we assume that $n \geq d$
  for the remainder of the paper.
The worst-case complexity of solving linear programs has traditionally been
  stated in terms of $n$, $d$, and $L$, where
  $L$ is commonly called the ``bit-length'' of the input linear program, but is rarely
  defined to actually be the number of bits necessary to specify the linear program.
For integer $A, \bb,\cc $, Khachiyan \cite{Khachiyan} and
   Karmarkar \cite{Karmarkar}
   defined $L$ to be some constant times 
\begin{multline*}
  \log (\mbox{largest absolute value of the determinant
   of any square sub-matrix of $A$})\\
 + 
  \log (\norm{\cc}_{\infty })
 +
  \log (\norm{\bb}_{\infty })
 +
  \log (n + d).
\end{multline*}
In the smoothed model, complexity estimates in terms
  of $L$ are quite pessimistic:
  even if one perturbs just the least significant digit
  of each entry of $A$, 
  the resulting $L$ value
  is at least some constant times $d$ with high probability.
Thus, in the smoothed model, our analysis of the complexity of
  interior point methods the replaces $L$, which is typically $\Omega(d)$,
  with $\log (nd /\sigma )$.

\subsection{Renegar's Condition Number}\label{ssec:RenegarsWork}

In~\cite{RenegarFunc,RenegarCond,RenegarPert},
  Renegar defined the condition number
  $C (A,\bb,\cc)$
  of a linear program and proved that an interior point algorithm
  whose complexity was
  $O (n^{3} \log (C (A,\bb,\cc) /\epsilon))$
  could solve a linear program to relative accuracy $\epsilon$, or
  determine that the program was infeasible or unbounded.

For a linear program in the canonical form (\ref{eqn:form1}),
  we follow Renegar~\cite{RenegarPert,RenegarCond,RenegarFunc} in
  defining the primal condition number, $C^{(1)}_{P} (A,\bb )$,
  of the program to be
  the normalized reciprocal of the distance to ill-posedness.
A program is ill-posed if the program can be made both feasible and
  infeasible by arbitrarily small changes to the pair $(A,\bb)$.
The distance to ill-posedness of the pair $(A,\bb)$ is the distance to the set
  of ill-posed programs under the Frobenius norm.
We similarly define the dual condition number, $C^{(1)}_{D} (A,\cc)$, to
  be the normalized reciprocal of the distance to ill-posedness of the
  dual program. The condition number, $C^{(1)} (A,\bb ,\cc )$,
  is the maximum of
  $C^{(1)}_{P} (A,\bb )$ and $C^{(1)}_{D} (A,\cc )$.

We can equivalently define the condition number without introducing
  the concept of ill-posedness.
For programs of form~(\ref{eqn:form1}), we define $C^{(1)}_{P} (A,\bb )$ by

\begin{definition}[Primal Condition Number]\label{def:conditionNumber}
\noindent
\begin{itemize}
\item [(a)] if $A \xx \leq \bb$ is feasible, then
\begin{eqnarray*}
\hspace{-.35in}
  C^{(1)}_{P} (A,\bb )
 =
\frac{
  \fnorm{A,\bb }}{
\sup \setof{\delta :
                \fnorm{\Delta A,  \Delta \bb} \leq \delta \mbox{ implies }
                (A + \Delta A) \xx \leq (\bb + \Delta \bb) \mbox{ is feasible}}},
\end{eqnarray*}

\item [(b)] if $A \xx \leq \bb$ is infeasible, then
\begin{eqnarray*}
\hspace{-.35in}
  C^{(1)}_{P} (A,\bb )
 =
\frac{
  \fnorm{A,\bb }
}{
 \sup \setof{\delta :
                \fnorm{\Delta A,  \Delta \bb} \leq \delta \mbox{ implies }
                (A + \Delta A) \xx \leq (\bb + \Delta \bb) \mbox{ is infeasible}}}.
\end{eqnarray*}
\end{itemize}
\end{definition}
It follows from the definition above that $C^{(1)}_{P} (A,\bb )\geq  1$.
We define the dual condition number, $C^{(1)}_{D}(A,\cc ) $, analogously.

To reader familiar with condition numbers in contexts
 outside of linear programming, the above definition may be surprising:
the condition numbers for numerous
  other problems are defined as the sensitivity
  of the output to perturbations in the input, and 
  are then often related to the distance to ill-posedness.
Renegar inverts this scheme
  by defining the condition number for linear programming to be the distance
  to ill-posedness, and then proving that the condition number does bound
  the sensitivity of the output to perturbations in the
  input~\cite{RenegarPert, RenegarCond}.

Any linear program may be expressed in  form~(\ref{eqn:form1});
 however, transformations among linear programming formulations
 do not in general
 (and commonly do not) preserve condition
  number~\cite{RenegarCond}.
We will therefore have to define different condition
  numbers for each normal form we consider.
For linear programs with canonical forms (\ref{eqn:form2}),
(\ref{eqn:form3}), and (\ref{eqn:form4})
  we define their condition numbers, $C^{(2)}(A,\bb ,\cc )$,
   $C^{(3)} (A,\bb ,\cc )$ and
  $C^{(4)} (A)$, analogously.
We follow the convention that $\bvec{0}$ is not considered
  a feasible solution to $(4)$.
Just as for $C^{(1)}_{P} (A,\bb )$, $C^{(i)}\geq 1$ for all $i$.

For linear programs given in form~(2),
Renegar~\cite{RenegarFunc,RenegarCond,RenegarPert}
  developed an initialization phase
  that returns a feasible point with
  initial optimality gap $R \leq O (n  C (A, \bb ,\cc ))$
  for a  linear program $(A,\bb ,\cc )$ or determines  that the program
  is infeasible or unbounded, in $O (n^{3} \log (C (A,\bb ,\cc )))$ operations.
By applying $O (\sqrt{n} \log (n C (A, \bb ,\cc )) / \epsilon )$
  iterations of a primal interior point method, for a total of
  $O (n^{3} \log (n C (A, \bb ,\cc )) / \epsilon )$
  arithmetic operations, Renegar proved:

\begin{theorem}[Renegar]\label{the:Renegar}
For any linear program of form~(2) specified by
  $(A,\bb ,\cc )$ and parameter $\epsilon$,
  Renegar's interior-point algorithm, in
  $O (n^{3} \log (nC (A,\bb ,\cc )/\epsilon ))$ operations,
  finds a feasible solution $\xx $ with optimality gap
  $\epsilon \fnorm{A,\bb ,\cc }$, or determines that the program is
  infeasible or unbounded.
\end{theorem}

Subsequently,
  algorithms with complexity logarithmic in the condition number
  were developed by 
  Vera \cite{Vera}
   for forms~(1) and~(3) and by 
  Cucker and  Pe\~{n}a \cite{CuckerPena} for form~(4).
The complexities of their algorithms are similar to that of Renegar's.
In \cite{FreundVera}, Freund and Vera give a unified approach which both
  efficiently estimates the condition number and solves the linear
  programs in any of these forms.

\subsection{Smoothed Analysis of Condition Number: Our Results}

In this paper, we 
  consider linear programming problems in which the data
  is subject to slight Gaussian perturbations.
Recall that the probability density function of
  a Gaussian random variable with mean $\orig{x}$ and variance $\sigma^{2}$
  is given by
\[
  \mu(x) = \frac{1}{\sigma \sqrt{2 \pi}} e^{- (x-\orig{x})^2/(2\sigma^2)}.
\]

A Gaussian perturbation of a vector $\orig{\xx }$ of
  variance $\sigma^2$ is a vector whose $i$th
  element is a Gaussian random variable of variance $\sigma^2$ and
  mean $\orig{x}_{i}$, and in which each element is independently chosen.
Thus, the probability density function of
  a $d$-dimensional Gaussian perturbation of
 $\orig{\xx }$ of variance $\sigma ^{2}$ is given by
\[
  \mu(\xx ) = \frac{1}{(\sigma \sqrt{2 \pi}  )^{d}}
   e^{-\norm{\xx - \orig{\xx }}^{2}/(2 \sigma ^{2})}.
\]
A Gaussian perturbation of a matrix may be defined similarly.

For each $(\orig{A},\orig{\bb},\orig{\cc })$ and
  $\sigma \geq 0$, 
  we let $\calN  ((\orig{A},\orig{\bb},\orig{\cc }),\sigma )$
  denote the distribution of Gaussian perturbations
  of  $(\orig{A},\orig{\bb},\orig{\cc })$ of variance $\sigma^{2}$,
  and we let $ (A,\bb ,\cc )\leftarrow \calN ((\orig{A},\orig{\bb},\orig{\cc }),\sigma )$
  indicate that $(A,\bb ,\cc )$ is drawn from the distribution
  $\calN ((\orig{A},\orig{\bb},\orig{\cc }),\sigma )$.

Our main result, which is proved in Section~\ref{sec:cond}, is

\begin{theorem}[Smoothed Complexity of Renegar's Condition
  Number]\label{thm:main}
For every $n$-by-$d$ matrix $\orig{A}$, $n$-vector $\orig{\bb}$ and
  $d$-vector $\orig{\cc }$
   such that $\fnorm{\orig{A}, \orig{\bb}, \orig{\cc}} \leq 1$,
  every $\sigma \leq 1/\sqrt{nd}$,
  and every $i \in \setof{1,2,3,4}$,
\begin{eqnarray*}
  \prob{A,\bb ,\cc }
       {C^{(i)} (A,\bb ,\cc )
            >
    \frac{2^{13}~ (n+1)^{2} (d+1)^{1.5} }{\delta \sigma^2}\left(\log
    \frac{2^{10}~(n+1)^2 (d+1)^{1.5}}{\delta \sigma^2} \right)^2
}
 & < & \delta,
\end{eqnarray*}
and
\begin{eqnarray*}
  \expec{(A,\bb ,\cc )
  \leftarrow \calN ((\orig{A},\orig{\bb},\orig{\cc }),\sigma ) }
       {\log C^{(i)}(A,\bb ,\cc )}
  & \leq &
    15 + 4.5 \log \frac{nd}{\sigma }.
\end{eqnarray*}
\end{theorem}

Theorem~\ref{thm:main} implies a bound on the smoothed complexity of Renegar's
algorithm as well as a bound on the smoothed complexity of the
interior point methods that were developed for the other canonical forms.
Note that in the statement of Theorem~\ref{thm:main}, we abuse the notation
  $C^{(4)} (A,\bb ,\cc )$ for
 $C^{(4)} (A)$.
The first bound of the theorem means that, with high
  probability, the condition number of a perturbed linear program
  is polynomial in $n$, $d$, and $1/\sigma $.

The following theorem follows immediately from Renegar's analysis
  (Theorem \ref{the:Renegar}) and the previous theorem.

\begin{theorem}[Smoothed Complexity of IPM]\label{thm:mainAlg}
Let $T ((A,\bb,\cc),\epsilon )$ be the time complexity of Renegar's interior
  point algorithm for finding $\epsilon$-accurate solutions
  of the linear program defined by $(A,\bb ,\cc )$ or
  determining that the program is infeasible or unbounded.
For every $n$-by-$d$ matrix $\orig{A}$, $n$-vector $\orig{\bb}$ and
  $d$-vector $\orig{\cc }$
   such that $\fnorm{\orig{A}, \orig{\bb}, \orig{\cc}} \leq 1$
  and every $\sigma \leq 1/\sqrt{nd}$,
\[
   \expec{(A,\bb ,\cc )
  \leftarrow \calN ((\orig{A},\orig{\bb},\orig{\cc }),\sigma )}
  {T ((A,\bb,\cc),\epsilon )} = O \left(n^{3} \log \left ( \frac{n}{\sigma \epsilon} \right)\right).
\]
\end{theorem}

In order to analyze Renegar's condition number for the primal and dual of each
  of the four canonical forms, we found it necessary to develop several extensions
  to the theory of condition numbers that may be of independent interest.
For example, Lemma~\ref{lem:maxMin} generalizes the geometric condition on distance
  to ill-posedness developed  in
  \cite{CheungCuckerNew} by incorporating an arbitrary non-pointed convex cone
  that is not subject to perturbation, and this generalization is necessary for the
  application of our techniques.
Additionally, Lemmas~\ref{lem:firstInfeasGeom},~\ref{lem:dualGeom}, and~\ref{lem:secondGeom}
  all provide geometric conditions on the distance to ill-posedness 
  whose import to us is on par with Lemma~\ref{lem:maxMin}.

\subsection{Organization of the Paper}

In our analysis, we divide the eight condition numbers $C_{P}^{(i)}$ and
  $C_{D}^{(i)}$, for $i\in \{1,2,3,4\}$, into
  two groups.
The first group includes $C_{P}^{(1)}$, $C_{P}^{(2)}$,
  $C_{D}^{(2)}$, $C_{D}^{(3)}$,
  and with some additional work, $C_{P}^{(4)}$.
The remaining condition numbers belong to the second group.
We will refer to a condition number from the first group as
  a {\em primal condition number} and a condition number from the
  second group as a {\em dual condition number}.

Section~\ref{sec:primal} is devoted to establishing a smoothed
  bound on the primal condition number.
%  on primal condition number of linear programs given in forms
%  (\ref{eqn:form1}), (\ref{eqn:form2}) and (\ref{eqn:form4}) as well as
%the dual condition number
%  of a linear program in Form (\ref{eqn:form2}) and (\ref{eqn:form3}).
We remark that the techniques used  in Section~\ref{sec:primal}
  do not critically depend upon $A$, $\bb$ and $\cc$ being
  Gaussian distributed, and similar theorems could be proved
  using slight modifications of our techniques if these
  were smoothly distributed within spheres or cubes.
It follows from the result of Section~\ref{sec:primal} alone
  that Theorem~\ref{thm:main} holds for linear program given in Form (\ref{eqn:form2}).

In Section~\ref{sec:dual}, we establish the smoothed bound
  on the dual condition number.
% for linear programs of the forms
%  (\ref{eqn:form1}) and (\ref{eqn:form4}) and primal condition number
%  of a linear program in form (\ref{eqn:form3}).
Our bounds in this section do critically make use of
  the Gaussian distribution on $A$, $\bb$ and $\cc$.

In Section~\ref{sec:cond}, we prove 
  Theorem~\ref{thm:main} using the smoothed bounds of the previous
  two sections.
We conclude the paper in Section~\ref{sec:discussion} with some
  open questions.

In the remainder of this Section, we review some of the previous work on
  smoothed analysis, some earlier results on the average-case analysis
  of interior-point algorithms, and lower bounds on the complexity of
  interior-point algorithms.

\subsection{Prior Smoothed Analyses of Linear Programming Algorithms}
In their paper introducing Smoothed Analysis~\cite{SpielmanTeng}, 
  Spielman and Teng 
  proved that the smoothed complexity of a two-phase
  shadow vertex simplex method was
  polynomial in $n$, $d$ and $1/\sigma $.
Shortly thereafter, Blum and Dunagan~\cite{BlumDunagan}
  performed a smoothed analysis of the perceptron algorithm
  for linear programming.
They showed that the probability the perceptron algorithm would
  take more than a polynomial in the input size times $k$
  steps was inversely proportional to $\sqrt{k}$.
Their analysis had the advantage of being significantly simpler
  than that of~\cite{SpielmanTeng}, and it is their
  analysis that we build upon in this work.
Blum and Dunagan's analysis used the fact that the number
  of steps taken by the perceptron algorithm can be
  bounded by the reciprocal of the
  ``wiggle room'' in its input, and the bulk
  of their analysis was a bound on the probability that this
  ``wiggle room'' was small.
The ``wiggle room'' turns out to be a condition number of
  the input to the perceptron algorithm.

\subsection{Prior Average-Case Analyses of Interior Point Algorithms}
There has been an enormous body of work on interior point algorithms,
  some of which has addressed their average-case complexity.
Anstreicher, Ji, Potra and Ye~\cite{AnstreicherJPY1,AnstreicherJPY2},
  have shown that under Todd's degenerate model for
  random linear programs~\cite{ToddModels},
  a homogeneous self-dual interior point method
  runs in $O (\sqrt{n} \log n)$ expected iterations.
Borgwardt and Huhn~\cite{BorgwardtHuhn} have obtained similar
  results under any spherically symmetric distribution.
The performance of other interior point methods on random inputs
  has been heuristically analyzed
  through ``one-step analyses'', but it is not clear
  that these analyses can be made rigorous~\cite{Nemirovsky,GonzagaTodd,MizunoAdaptive}.

\subsection{Lower Bounds for Interior Point Algorithms}

The best known lower bound on the complexity of interior
  point methods is $\Omega (n^{1/3})$ iterations
  due to Todd~\cite{ToddLower} and Todd and Ye~\cite{ToddYeLower}.
However, the programs for which these lower bounds hold
  are very ill-conditioned.
There are no known bounds of the form $\Omega (n^{\epsilon })$
  for well-conditioned linear programs.
It would be interesting to know whether such a lower bound
  can be proved for a well-conditioned program, or
  whether interior point algorithms always require fewer iterations    
  when their input is well-conditioned.

\subsection{Notation and Basic Geometric Definitions}\label{ssec:notation}

Throughout this paper we use the following notational conventions. The material up
to this point has obeyed these conventions.

\begin{itemize}
\item lower case letters such as $a$ and $\alpha $ denote scalars,

\item bold lower case letters such as $\aa$ and $\bb$ denote vectors,
      and for a vector $\aa $, $a_{i}$ denotes the $i$th entry of $\aa$.

\item capital letters such as $A$ denote matrices, and

\item bold capital letters such as $\CC $ denote convex sets.
\end{itemize}

If $\vp{\aa}{1}{n}$ are vectors, we let
  $[\vp{\aa}{1}{n}]$ denote the matrix whose rows
  are the $\aa _{i}$s.
For a vector $\aa$, we let $\norm{\aa}$ denote the standard
  Euclidean norm of the vector.
We will make frequent use of the Frobenius norm of a matrix,
  $\fnorm{A}$, which is the square root of the sum of squares
  of the entries in the matrix.
We extend this notation to let $\fnorm{A, \vp{\xx}{1}{k}}$
   denote the
  square root of the sum of squares of the entries in $A$
  and in $\vp{\xx}{1}{k}$.
Different choices of norm are possible;
  we use the Frobenius norm throughout this paper.
The following proposition relates several common choices of norm:

\begin{proposition}[Choice of norm]\label{pro:norms}
For an $n$-by-$d$ matrix $A$,
\begin{eqnarray*}
  \frac{\fnorm{A}}{\sqrt{dn}}  & \leq ~ \norm{A}_{\infty } ~ \leq &\fnorm{A},
 \mbox{ and }\\
  \frac{\fnorm{A}}{\sqrt{d}}  & \leq ~ \norm{A}_{OP} ~ \leq &\fnorm{A},
\end{eqnarray*}
where $\norm{A}_{OP}$ denotes the operator norm of $A$, $\max_{x
\neq 0} \frac{\norm{Ax}}{\norm{x}}$.
\end{proposition}

We let $\log$ denote the logarithm to base 2 and $\ln$ denote the logarithm to base $e$.

We also make use of the following
  geometric definitions:

\begin{definition}[Ray]\label{def:ray}
For a vector $\pp$, let $\ray{\pp}$ denote
  $\setof{\alpha \pp : \alpha > 0}$.
\end{definition}

\begin{definition}[Non-pointed convex cone]\label{def:cone}
A non-pointed convex cone is a convex set $\CC$ such that for all
$\xx \in \CC$
  and all $\alpha > 0$, $\alpha \xx \in \CC$, and
  there exists a vector
  $\tt$ such that $\form{\tt}{\xx} < 0$ for all $\xx \in \CC$.
\end{definition}

\begin{definition}[Positive half-space]\label{def:pos}
For a vector $\aa$ we let $ \pos{\aa}$ denote the
  half-space of points with non-negative inner product with $\aa$.
\end{definition}

For example, $\Reals{d}$ and $\pos{\xx}$ are not non-pointed convex
cones,
  while $\setof{\xx : \xx _{0} > 0}$
  and $\ray{\pp}$ are non-pointed convex cones.
Note that a non-pointed convex cone cannot contain the origin. All
of the cones that we introduce through the process of
homogenization are non-pointed convex cones.

These definitions enable us to express the feasible $\xx$
  for the linear program
\[
  A \xx \geq \bvec{0} \mbox{ and } \xx \in \CC
\]
as
\[
  \xx  \in \CC \intersect \bigcap_{i=1}^{n} \pos{\aa _{i}},
\]
where $\vp{\aa}{1}{n}$ are the rows of $A$. Throughout this paper,
we will call a set
 \emph{feasible} if it is non-empty, and
 \emph{infeasible} if it is empty.
Thus, we say that  the set $\CC \intersect \bigcap_{i=1}^{n} \pos{\aa
_{i}}$
  is feasible if the corresponding
  linear program is feasible.

% Local Variables: ***
% TeX-master:"journalipm.tex" ***
% End: ***

\section{Primal Condition Number}\label{sec:primal}

\setcounter{theorem}{0}

In this section we show that the
  smoothed value of the
  primal condition numbers
  is polynomial in $n$, $d$, and $1/\sigma$
  with polynomially high probability.
As in the work of Pe{\~{n}}a~\cite{JavierUnderstanding},
  we unify this study by transforming each
  canonical form to conic form.

The primal program of form~(1) can be put into conic form with the
  introduction of the homogenizing variable $x_{0}$.
Setting $\CC = \setof{(\xx ,x_{0}) : x_{0} > 0}$,
 the homogenized primal program of form~(1) is
\[
   [-A, \bb] (\xx , x_{0}) \geq \bvec{0}, ~ (\xx ,x_{0}) \in \CC.
\]
By
  setting $\CC = \setof{(\xx ,x_{0}) : x_{0} > 0\ and \ \xx \geq
  \bvec{0}}$, one can similarly homogenize
  the primal program of form~(2).
The dual programs of form~(2) and form~(3) can be homogenized
  by setting $\CC = \setof{(\yy ,y_{0}) : y_{0} > 0}$ and
  $\CC = \setof{(\yy ,y_{0}) : y_{0} > 0\ and \ \yy \geq
  \bvec{0}}$, respectively, and considering the program
\[
   [-A^{T}, \cc] (\yy , y_{0}) \geq \bvec{0}, ~ (\yy ,y_{0}) \in \CC.
\]
We will comment on $C_{P}^{(4)}$ below. 
Note that in each of these homogenized programs,
  the variables lie in a non-pointed convex cone.

Pe{\~{n}}a~\cite{JavierUnderstanding} proves:

\begin{fact}[Preserving feasibility]\label{fac:iff}
Each of the homogenized programs is feasible if and only if its
  original program is feasible.
\end{fact}

In Section \ref{sub:conic},
  we extend the notion of distance to ill-posedness and condition number
  to conic linear programs and note that
  the transformation by homogenization does not alter the distance
  to ill-posedness.
The rest of the section will be devoted to analyzing
  the condition number of the conic program, and this will imply the
  bound on the condition number of the original program.

\subsection{Linear Programs in Conic Form and Basic Convex Probability Theory}
\label{sub:conic}
The feasibility problem for a conic linear program can be written:
\[
  \mbox{find $\xx$ such that }  A\xx \geq \bvec{0}, \xx \in \CC,
\]
where $\CC $ is a non-pointed convex cone in $\Reals{d}$ and $A$
is an $n$-by-$d$ matrix.
Note that because $\CC $ is a non-pointed convex cone,
  $\bvec{0}$ cannot be a feasible solution
  of this program.
The following definition generalizes distance to
  ill-posedness by explicitly taking into account the non-pointed convex
  cone, $\CC$.

%The following is a generalization of the distance to ill-posedness
%  that we will use throughout this section.

\begin{definition}[Generalized distance to ill-posedness]\label{def:rho}
For a non-pointed convex cone,  $\CC$, that is not subject to
perturbation, and a matrix, $A$,
  we define $\rho (A, \CC )$ by
\begin{itemize}
\item [a.]
  if $A \xx \geq \bvec{0}, ~ \xx \in C$ is feasible, then
\[
 \rho(A , \CC )
 = \sup \setof{\epsilon : \fnorm{\Delta A} < \epsilon \mbox{ implies }
          (A + \Delta A) \xx \geq \bvec{0}, ~ \xx \in \CC  \mbox{ is
feasible}};
\]
\item [b.]
  if $A \xx \geq \bvec{0}, ~ \xx \in C$ is infeasible, then
\[
 \rho(A , \CC )
 = \sup \setof{\epsilon : \fnorm{\Delta A} < \epsilon \mbox{ implies }
          (A + \Delta A) \xx \geq \bvec{0}, ~ \xx \in \CC  \mbox{ is infeasible}}.
\]
\end{itemize}
\end{definition}
We note that this definition makes sense even when $A$ is a row
  vector.
In this case,
  $\rho(\aa , \CC)$ measures the distance to ill-posedness when we
  only allow perturbation to $\aa $.
Even though transformations among linear programming
  formulations in general do not preserve condition number,
  Pe{\~{n}}a~\cite{JavierUnderstanding} has proved that
  homogenization does not alter the distance to ill-posedness.
For convenience, we will state the lemma
  for form~(1), and note that similar statements hold for $C_{P}^{(2)}$,
  $C_{D}^{(2)}$, and $C_{D}^{(3)}$.

\begin{lemma}[Preserving the condition number]\label{lem:homogenizing}
Let
\[
\max~ \cc^{T}\xx \quad  s.t. \quad A\xx \leq \bb
\]
be a linear   program.
Let $\CC = \setof{(\xx ,x_{0}) : x_{0} > 0}$.
Then, $C_{P}^{(1)} (A,\bb ) =   \fnorm{A,\bb } / \rho ([-A,\bb],\CC ) $.
\end{lemma}

The primal program of form~(4) is not quite in conic form;
  to handle it, we need the following definition.

\begin{definition}[Pointed generalized primal distance to ill-posedness]\label{def:rho2}
For a convex cone that is {\em not non-pointed}, $\CC$, and a
matrix, $A$,
  we define $\rho (A, \CC )$ by
\begin{itemize}
\item [a.]
  if $A \xx \geq 0, ~ \xx \not = \bvec{0}, ~ \xx \in C$ is feasible, then
\[
 \rho(A , \CC )
 = \sup \setof{\epsilon : \fnorm{\Delta A} < \epsilon \mbox{ implies }
          (A + \Delta A) \xx \geq \bvec{0}, ~ \xx \not = \bvec{0},
            ~ \xx \in \CC  \mbox{ is feasible}}
\]
\item [b.]
  if $A \xx \geq 0, ~ \xx \not = \bvec{0}, ~ \xx \in C$ is infeasible, then
\[
 \rho(A , \CC )
 = \sup \setof{\epsilon : \fnorm{\Delta A} < \epsilon \mbox{ implies }
          (A + \Delta A) \xx \geq \bvec{0}, ~ \xx \not = \bvec{0},
             ~ \xx \in \CC  \mbox{ is infeasible}}
\]
\end{itemize}
\end{definition}
This definition would allow us to prove the analogs of
  Lemmas~\ref{lem:primal} and ~\ref{lem:logcond} for primal programs of form~(4).
We omit the details of this variation on the arguments
  in the interest of simplicity.

The following two Lemmas are the main result of this section.
To see how they may be applied, we note that
  a simple union bound over $C^{(2)}_P$ and
  $C^{(2)}_D$ using Lemma~\ref{lem:primal}
  yields Theorem~\ref{thm:main} for form (\ref{eqn:form2}).

\begin{lemma}[Condition number is likely polynomial]\label{lem:primal}
For any non-pointed convex cone $\CC $
  and a matrix $\orig{A}$ satisfying
  $\fnorm{\orig{A}} \leq 1$, for $\sigma \leq 1/\sqrt{nd}$,
\begin{eqnarray*}
\prob{A \leftarrow \calN (\orig{A},\sigma)}
     {\frac{\fnorm{A}}{\rho (A, \CC )} ~\geq~
 \frac{2^{12} n^{2} d^{1.5}
 }{
 \delta \sigma^2
 }
    \log^2
    \left(\frac{2^9 n^2 d^{1.5}}{\delta \sigma^2}\right)
} & \leq & \delta.
\end{eqnarray*}
\end{lemma}

\begin{lemma}[Smoothed analysis of log of primal condition number]\label{lem:logcond}
For any non-pointed convex cone $\CC $
  and a matrix $\orig{A}$ satisfying
  $\fnorm{\orig{A}} \leq 1$, for $\sigma \leq 1/\sqrt{nd}$,
\begin{eqnarray*}
\expec{A \leftarrow \calN (\orig{A},\sigma)}
     {\log \frac{\fnorm{A}}{\rho (A, \CC )}}
 \leq 14 + 4.5\log \frac{nd}{\sigma}.
\end{eqnarray*}
\end{lemma}

We will prove Lemma \ref{lem:primal}
   by separately considering the cases in which the program is
  feasible and infeasible.
%The analysis of $C_P$ will proceed as follows: we consider the cases
%  that the program is feasible and infeasible separately.
In Section~\ref{ssec:primFeas},
  we show that it is unlikely that a program
  is feasible and yet can be made infeasible
  by a small change to its constraints (Lemma~\ref{lem:firstFeas}).
In Section~\ref{ssec:primInfeas}, we show that
  it is unlikely that a program is infeasible
  and yet can be made feasible by a small change
  to its constraints (Lemma~\ref{lem:firstInfeas}).
In Section~\ref{ssec:primBoth}, we combine these results
  to show that the
  primal condition number is polynomial with high probability (Lemma~\ref{lem:primal}).
In Section~\ref{ssec:logexp} we prove Lemma \ref{lem:logcond}.

The thread of argument in these sections
  consists of a geometric characterization
  of those programs with poor condition number, followed by a
  probabilistic argument demonstrating that this characterization
  is rarely satisfied.
Throughout the proofs in this section, $\CC $ will always refer
  to the original non-pointed cone, and a subscripted $\CC$ (\textit{e.g.}, $\CC_{0}$) will
  refer to a modification of this cone.

The key probabilistic tool used in the analysis is
  Lemma \ref{lem:small}, which we will derive from the following
  result of~\cite{Ball}.
A slightly
  weaker version of this lemma was proved in \cite{BlumDunagan}, and also
  in \cite{Obscure2}.

\begin{theorem}[Ball~\cite{Ball}]\label{thm:Ball}
Let $\KK$ be a convex body in $\Reals{d}$ and let
  $\mu$ be the density function of a $\calN (\bvec{0},\sigma)$
  Gaussian random variable.
Then,
\[
  \int_{\partial \KK} \mu \leq 4 d^{1/4}.
\]
\end{theorem}

\begin{lemma}[$\epsilon$-Boundaries are likely to be missed]\label{lem:small}
Let $\KK$ be an arbitrary convex body in $\Reals{d}$,
  and let $\bdry{\KK,\epsilon}$
  denote the
  $\epsilon$-boundary of $\KK$; that is,
\[
  \bdry{\KK,\epsilon}
  = \setof{\xx  : \exists \xx ' \in \partial \KK,
  \norm{\xx -\xx '} \leq \epsilon}
\]
For any $\orig{\xx } \in \Reals{d}$,
\begin{align*}
&  \prob{\xx \leftarrow \calN (\orig{\xx}, \sigma )}
    {\xx  \in \bdry{\KK,\epsilon} \setminus \KK }
 \leq
  \frac{4 \epsilon d^{1/4}}{\sigma},
 && \mbox{{(outside boundary) }}\\
&  \prob{\xx \leftarrow \calN (\orig{\xx}, \sigma )}
   {\xx  \in \bdry{\KK,\epsilon} \cap  \KK }
 \leq
  \frac{4 \epsilon d^{1/4}}{\sigma}
 && \mbox{{(inside boundary) }}
\end{align*}
\end{lemma}
\begin{proof}
We derive the result assuming $\sigma = 1$.
The result for general $\sigma$ follows by scaling.
%The result for general $\sigma$ can be obtained by contracting
%  the picture by a factor of $\sigma$.

Let $\mu $ denote the density according to which $\xx$
  is distributed.
To derive the first inequality, we let $\KK_{\epsilon}$ denote
  the points of distance at most $\epsilon$ from $\KK$,
  and observe that $\KK_{\epsilon}$ is convex.

Integrating by shells, we obtain
\begin{align*}
  \prob{}{\xx  \in \bdry{\KK,\epsilon} \setminus \KK }
& \leq
  \int_{t = 0}^{\epsilon }
  \int_{\partial \KK_{t}} \mu \\
& \leq
   \epsilon 4 d^{1/4},
\end{align*}
by Theorem~\ref{thm:Ball}.

We similarly derive the second inequality by defining
  $\KK^{\epsilon}$ to be the set of points inside $\KK$
  of distance at least $\epsilon$ from the boundary of $\KK$
  and observing that $\KK^{\epsilon}$ is convex for any $\epsilon $.
\end{proof}

In this section and the next, we use the following consequence
  of Lemma~\ref{lem:small} repeatedly.

\begin{lemma}[Feasible likely quite feasible, single constraint]\label{lem:wide-one}
Let $\CC_{0}$ be any convex cone in $\Reals{d}$ and,
  for any $\orig{\aa }\in\Reals{d}$, let $\aa$
  be a Gaussian perturbation of $\orig{\aa}$ of
  variance $\sigma^{2}$.
Then,
\begin{align*}
 \prob{\aa}
      {\CC_{0} \cap \pos{\aa} \mbox{ is feasible and } \rho (\aa, \CC_{0}) \leq \epsilon }
& \leq
  \frac{4 \epsilon d^{1/4}}{\sigma},
 \mbox{ and}\\
 \prob{\aa}
      {\CC_{0} \cap \pos{\aa} \mbox{ is infeasible and } \rho (\aa, \CC_{0}) \leq \epsilon }
& \leq
 \frac{4 \epsilon d^{1/4}}{\sigma}.
\end{align*}
\end{lemma}

\begin{proof}
Let $\KK$ be the set of $\aa $ for which
  $\CC_{0} \intersect \pos{\aa }$ is infeasible.
Observe that $\rho (\aa , \CC_{0})$ is exactly the distance from
  $\aa$ to the boundary of $\KK$.
Since $\KK$ is a convex cone,
  the first inequality follows from the first inequality (the outside
  boundary inequality)
  of Lemma~\ref{lem:small}, which tells us that the probability
  that $\aa $
  has distance at most $\epsilon$ to the boundary of $\KK$
  and is outside $\KK$
  is at most
  $\frac{4 \epsilon d^{1/4}}{\sigma}$.
The second inequality similarly follows from the second inequality
  (the inside boundary inequality)
  of Lemma~\ref{lem:small}.
\end{proof}

\subsection{Primal condition number, feasible case}\label{ssec:primFeas}

In this subsection, we analyze the primal condition number
  in the feasible case and prove:

\begin{lemma}[Feasible is likely quite feasible, all constraints]\label{lem:firstFeas}
Let $\CC$ be a non-pointed convex cone in $\Reals{d}$ and let
  $\orig{A}$ be any $n$-by-$d$ matrix.
Then for any  $\sigma \geq 0$,
\[
  \prob{A \leftarrow \calN (\orig{A},\sigma)}{
     \left( A \xx \geq \bvec{0}, ~ \xx \in \CC
          \mbox{ is feasible} \right)
     \mbox{ and }
     \left( \rho (A, \CC) \leq \epsilon \right)}
  \leq
   \frac{4 \epsilon n d^{5/4}}{\sigma}.
\]
\end{lemma}

To prove Lemma \ref{lem:firstFeas}, we first
  establish a necessary geometric condition for $\rho$ to be
  small.
This condition is stated and proved in Lemma~\ref{lem:maxMin}.
In Lemma~\ref{lem:maxMinBound}, we apply Helly's Theorem~\cite{Helly} to
  simplify this geometric condition, expressing it in terms of
  the minimum of $\rho $ over individual constraints. This allows us to
  use Lemma~\ref{lem:wide-one}
  to establish Lemma~\ref{lem:maxMinProb}, which shows that
  this geometric condition is unlikely to be met.
Lemma \ref{lem:firstFeas} is then a corollary of
  Lemmas~\ref{lem:maxMinProb}~and~\ref{lem:maxMin}.

We remark that a result similar to Lemma~\ref{lem:maxMin}
  appears in~\cite{CheungCuckerNew}.

%All of the remaining lemmas of this section except the last are
%  used to establish a necessary geometric condition for $\rho$ to be
%  small.
%All the lemmas of this section but the last are used
%  to establish a necessary
%  geometric condition for $\rho $ to be small.
%The last lemma, Lemma \ref{lem:maxMinProb}, uses
%  Lemma~\ref{lem:wide-one} to
%  show that this geometric condition is unlikely to be met.
%We then prove the main lemma of the subsection, Lemma \ref{lem:firstFeas}.

\begin{lemma}[Bounding $\rho$ by a max of min of inner products]\label{lem:maxMin}
Let $\CC$ be a non-pointed convex cone
  and let $\vs{\aa}{1}{n}$ be vectors in $\Reals{d}$
  for which
  $\CC \cap \bigcap_{i} \pos{\aa _{i}}$
  is feasible.
Then
\[
  \rho ([\vs{\aa}{1}{n}], \CC)
\geq
%DAS
 \max_{\substack{\pp \in \CC \cap \bigcap_{i=1}^{n} \pos{\aa_{i}}  \\
         \norm{\pp} = 1}}
%SHT
%  \max_{\pp \in \CC \cap \bigcap_{i=1}^{n} \pos{\aa_{i}}: \norm{\pp} = 1}
  \min_{i} \aa_{i}^{T} \pp.
\]
\end{lemma}

\begin{proof} Lemma~\ref{lem:maxMin} follows directly
   from Lemmas~\ref{lem:rhoRay}, \ref{lem:maxOverRays},
   and \ref{lem:rhoSetRay} below.
These three lemmas develop a characterization of $\rho$, the
  distance to ill-posedness, in the feasible case.
\end{proof}

\begin{lemma}[Lower bounding $\rho$ by rays]\label{lem:maxOverRays}
Under the conditions of Lemma~\ref{lem:maxMin},
\[
  \rho ([\vs{\aa}{1}{n}], \CC )
\geq
  \max _{\pp \in \CC \cap \bigcap_{i} \pos{\aa _{i}}}
  \rho ([\vs{\aa}{1}{n}], \ray{\pp }) .
\]
\end{lemma}
\begin{proof}
Let $\vs{\Delta \aa}{1}{n}$ be such that
  $\CC  \cap \bigcap_{i} \pos{\Delta \aa _{i} + \aa _{i}}$
  is infeasible.
Then, for all $\pp \in \CC \cap \bigcap_{i} \pos{\aa _{i}}$,
  $\ray{\pp } \cap \bigcap_{i} \pos{\Delta \aa _{i} + \aa _{i}}$
  is also infeasible.
\end{proof}

\begin{lemma}[$\rho $ of a ray as a min over constraints]\label{lem:rhoSetRay}
For every set of vectors $\vs{\aa}{1}{n}$ and $\pp$ such that
 $\ray{\pp} \intersect \bigcap_{i} \pos{\aa_{i} }$ is feasible,
\[
  \rho ([\vs{\aa}{1}{n}], \ray{\pp})
  =
  \min_{i} \rho (\aa_{i}, \ray{\pp}).
\]
\end{lemma}
\begin{proof}
Observe that
  $\ray{\pp} \intersect \bigcap_{i} \pos{\aa_{i} + \Delta \aa_{i}}$
  is feasible if and only if
  $\ray{\pp} \intersect \pos{\aa_{i} + \Delta \aa_{i}}$
  is feasible for all $i$.
\end{proof}

\begin{lemma}[$\rho $ of a ray and single constraint as an inner product]\label{lem:rhoRay}
For every vector $\aa$ and every unit vector $\pp$,
\[
   \rho (\aa ,\ray{\pp}) = \abs{\form{\aa}{\pp}}
\]
\end{lemma}
\begin{proof}
If $\aa ^{T} \pp = 0$, then $\rho (\aa , \ray{\pp} ) = 0$.
If $\aa ^{T} \pp \not = 0$, then
  $\ray{\pp} \intersect \pos{\aa }$ is feasible if and only if
  $\ray{-\pp} \intersect \pos{\aa }$ is infeasible; so, it suffices to consider
  the case where $\ray{\pp} \intersect \pos{\aa }$ is feasible.
So, we assume $\form{\aa}{\pp} > 0$, in which case
 $\ray{\pp} \intersect \pos{\aa }$ is feasible.
We first prove that
  $\rho (\aa ,\ray{ \pp}) \geq  \form{\aa}{\pp}$.
For every vector $\Delta \aa$ of norm at most $\form{\aa}{\pp}$,
  we have
\[
   \form{(\aa + \Delta \aa)}{\pp}
 =
   \form{\aa}{\pp} + \form{\Delta \aa}{\pp}
 \geq
   \form{\aa}{\pp} - \norm{\Delta \aa}
 \geq
   0.
\]
That is, $\pp  \in \pos{\aa + \Delta \aa }$.
As this holds for every $\Delta \aa$ of norm at most
  $\form{\aa}{\pp}$, we have
    $\rho (\aa ,\ray{ \pp}) \geq  \form{\aa}{\pp}$.

To show that $\rho (\aa ,\ray{ \pp}) \leq  \form{\aa}{\pp}$,
  note that setting
  $\Delta \aa = - (\epsilon + \form{\aa}{\pp}) \pp$, for any $\epsilon > 0$, yields
\[
   \form{(\aa + \Delta \aa)}{\pp}
 =
  \form{\aa}{\pp} + \form{\Delta \aa}{\pp}
 =
  \form{\aa}{\pp} - \form{(\epsilon + \form{\aa}{\pp}) \pp}{\pp}
 =
  \form{\aa}{\pp} - (\epsilon + \form{\aa}{\pp})
  = -\epsilon;
\]
so, $\ray{\pp} \intersect \pos{\aa + \Delta \aa }$ is infeasible.
As this holds for every $\epsilon > 0$,
  $\rho (\aa ,\ray{ \pp}) \leq  \form{\aa}{\pp}$.
\end{proof}

\begin{lemma}[Bounding the max of min of inner products]\label{lem:maxMinBound}
Let $\CC$ be a non-pointed convex cone
  and let $\vs{\aa}{1}{n}$ be vectors in $\Reals{d}$
  for which
  $\CC \cap \bigcap_{i} \pos{\aa _{i}}$
  is feasible.
Then
\[
  \max_{\substack{\pp \in \CC \cap \bigcap_{i=1}^{n} \pos{\aa_{i}} \\
         \norm{\pp} = 1}}
  \min_{i} \aa_{i}^{T} \pp
\geq
  \min_{i} \rho \left(
                 \aa _{i}, \CC \cap \bigcap_{j \not = i} \pos{\aa_{j}}
               \right) \Bigg/ d.
\]
\end{lemma}

We will derive Lemma~\ref{lem:maxMinBound} from Lemmas~\ref{lem:rhoCone} 
  and~\ref{lem:helly}, which we now state and prove.

\begin{lemma}[Quite feasible region implies
quite feasible point, single constraint]\label{lem:rhoCone} For
every $\aa$ and every non-pointed convex cone $\CC_{0}$
  for which $\CC_{0} \intersect \pos{\aa }$ is feasible,
\[
  \rho (\aa ,\CC_{0})
 =
  \max_{\substack{\pp \in \CC_{0} \cap \pos{\aa } \\
                  \norm{\pp} = 1}}
  \form{\aa}{\pp}.
\]
\end{lemma}
\begin{proof}
The ``$\geq$'' direction
  follows from Lemmas~\ref{lem:maxOverRays} and~\ref{lem:rhoRay};
  so, we
  concentrate on showing
\[
  \rho (\aa ,\CC_{0})
\leq
  \max_{\substack{\pp \in \CC_{0} \cap \pos{\aa } \\
                  \norm{\pp} = 1}}
  \form{\aa}{\pp}.
\]
We recall that, as $\CC_{0}$ is non-pointed, there exists a vector $\tt$ such that
  $\form{\tt}{\xx} < 0$ for all $\xx \in \CC_{0}$.
We now divide the proof into two cases depending on whether
   $\aa \in \CC _{0}$.

If $\aa \in \CC _{0}$, then we let $\pp = \aa / \norm{\aa }$.
It is easy to verify that
\[
\aa ^{T} \pp  =  \norm{\aa } =
  \max _{\norm{\pp } = 1} \form{\aa }{\pp } =
   \max_{\substack{\pp \in \CC_{0} \cap \pos{\aa } \\
                  \norm{\pp} = 1}} \form{\aa}{\pp}.
\]
Moreover,
  $\CC _{0} \intersect \pos{\aa - (\aa + \epsilon \tt)}$
  is infeasible  for every $\epsilon > 0$.
So, $\rho (\aa , \CC _{0}) \leq \norm{\aa}$.

If $\aa \not \in \CC _{0}$,
  let $\qq$ be the point of $\CC_{0}$ that is closest to $\aa$.
As $\CC_{0} \intersect \pos{\aa }$ is feasible, $\qq$
  lies inside $\pos{\aa }$ and
  is not the origin.
Let $\pp = \qq / \norm{\qq}$.
As $\CC _{0}$ is a cone, $\qq$ is perpendicular to
  $\aa - \qq$.
Thus, the distance from $\aa$ to $\qq$ is
  $\sqrt{\norm{\aa}^{2} - \norm{\qq}^{2}} =
  \sqrt{\norm{\aa}^{2} - (\form{\aa}{\pp})^{2}}$,
  as $\form{\aa}{\pp } = \norm{\qq}$.
Conversely, for any unit vector $\rr \in \CC _{0}$,
  the distance from $\ray{\rr }$ to $\aa $ is
  $\sqrt{\norm{\aa}^{2} - (\form{\aa}{\rr })^{2}}$.
Thus, the unit vector $\rr \in \CC _{0}$ 
  maximizing $\form{\aa}{\rr}$
  must be $\pp $.

As $\CC_{0}$ is convex, there is a plane through $\qq$
  separating $\CC_{0}$ from $\aa$ and perpendicular to the
  line segment $\aa - \qq$, and thus
  $\rho (\aa ,\CC _{0}) \leq \norm{\qq} =\form{\aa}{\pp}$.

%Thus, every point of $\CC_{0}$ has inner product at most zero
%  with the vector $\aa - \qq$.
%So,
%  $\CC_{0} \intersect \pos{\aa - \qq + \epsilon \tt }$ is infeasible
%  for every $\epsilon > 0$,
%  from which we may conclude
%  $\rho (\aa ,\CC _{0}) \leq \norm{\qq}$.
%To conclude the proof, we note that $\norm{\qq} = \form{\aa}{\pp}$.
\end{proof}

\begin{lemma}[Quite feasible individually implies quite feasible collectively]
\label{lem:helly}
Let $\CC_{0}$ be a non-pointed convex cone and let $\vs{\aa}{1}{n}$
  be vectors in $\Reals{d}$.
If there exist
  unit vectors $\vs{\pp}{1}{n} \in \CC_0$,
  such that
\begin{eqnarray*}
\form{\aa_{i}}{\pp_{i}} & \geq & \epsilon, \mbox{ for all $i$, and}\\
\form{\aa_{i}}{\pp_{j}} & \geq & 0, \mbox{ for all $i$ and $j$,}
\end{eqnarray*}
then there exists a unit vector $\pp \in \CC _{0}$ such that
\[
   \form{\aa_{i}}{\pp} \geq \epsilon / d, \mbox{ for all $i$.}
\]
\end{lemma}
\begin{proof}
We prove this using Helly's Theorem~\cite{Helly} which says that
  if a collection of convex sets in $\Reals{d}$ has the property
  that every subcollection of
  $d+1$ of the sets has a common point, then the entire
  collection has a common point.
Let
\[
  \SS _{i}  = \{\xx \in \CC_0 :
       \aa_i^T\xx/\norm{\xx} \geq \epsilon/d \}.
\]

We begin by proving that every $d$ of the $\SS _{i}$s
  contain a point in common.
Without loss of generality, we consider $\vs{\SS }{1}{d}$.
Let $\pp =  \sum_{i=1}^{d} \pp_{i} / d$.
Then, for each $1 \leq j \leq d$,
\[
  \form{\aa_{j}}{\pp}
 =
  \form{\aa_{j}}{\left( \sum_{i=1}^{d} \pp_{i} / d \right)}
 \geq
  \form{\aa_{j}}{\left( \pp_{j}/d \right)}
 \geq
   \epsilon / d.
\]
As $\pp$ has norm at most one, 
  $\form{\aa_{j}}{\pp / \norm{\pp}} \geq \form{\aa_{j}}{\pp}$,
  so $\pp$ is contained in each of $\vs{S}{1}{d}$.

As $\CC_0$ is non-pointed, there exists $\tt$ such that $\tt^T\xx < 0,
   ~\forall \xx \in \CC_0$.
Let $\SS'_{i} = \SS _{i} \bigcap \{\xx : \tt^T\xx = -1 \}$.
Then, $\xx \in \SS _{i}$ implies $-\xx  / \tt ^{T} \xx \in \SS'_{i}$.
So, every $d$ of the $\SS'_{i}$ have a point in common.
As these are convex sets lying in a $d-1$ dimensional space,
  Helly's Theorem tells us that there exists a point $\pp $
  that lies within all of the $\SS '_{i}$s.
As $\SS '_{i} \subset \SS _{i}$, this point lies inside all the
  $\SS _{i}$s.
\end{proof}

\begin{proof}[Proof of Lemma~\ref{lem:maxMinBound}]
For each $i$, we apply Lemma~\ref{lem:rhoCone}, 
  to the vector $\aa _{i}$ and the cone 
  $\CC \cap \bigcap_{j \not = i} \pos{\aa _{j}}$ to
  find a unit vector
  $\pp _{i} \in \CC \cap \bigcap_{j=1}^{n} \pos{\aa _{j}}$
  such that
\[
  \form{\pp _{i}}{\aa _{i}}
=
 \rho \left(\aa _{i}, \CC \cap \bigcap_{j \not = i} \pos{\aa _{j}} \right).
\]
As $\pp _{i} \in \CC \cap \bigcap_{j} \pos{\aa _{j}}$, we also have
\[
  \form{\pp _{i}}{\aa _{j}}
\geq
  0
\]
for all $j$.
Applying Lemma~\ref{lem:helly}, we find a unit vector
  $\pp \in \CC \cap \bigcap_{j=1}^{n} \pos{\aa _{j}}$
  satisfying
\[
  \form{\aa _{i}}{\pp }
 \geq \min _{i}
  \rho \left(\aa _{i}, \CC \cap \bigcap_{j \not = i} \pos{\aa _{j}}
\right) \Bigg/ d,
\]
for all $i$.
\end{proof}

\begin{lemma}[Max of min of inner products is likely large]\label{lem:maxMinProb}
Let $\CC$ be a non-pointed convex cone in $\Reals{d}$ and let
 $\vs{\orig{\aa}}{1}{n}$ be vectors in $\Reals{d}$.
 %such that $\fnorm{[\vs{\orig{\aa}}{1}{n}]} \leq  1$.
Let $\vs{\aa}{1}{n}$ be Gaussian perturbations of
  $\vs{\orig{\aa}}{1}{n}$  of variance  $\sigma ^{2}$.
Then,
\[
  \prob{}{
   C \cap \bigcap_{i}\pos{\aa _{i}} \mbox{ is feasible and }
%DAS
  \max_{\substack{\pp \in \CC \cap \bigcap_{i=1}^{n} \pos{\aa_{i}} \\
         \norm{\pp} = 1}}
%SHT
%  \max_{\pp \in \CC \cap \bigcap_{i=1}^{n} \pos{\aa_{i}} : \norm{\pp} = 1}
  \min_{i} \aa_{i}^{T} \pp < \epsilon
  }
  \leq
  \frac{4 \epsilon n d^{5/4}}{\sigma}.
\]
\end{lemma}
\begin{proof}
By Lemma~\ref{lem:maxMinBound},
\begin{multline*}
  \prob{}{
   C \cap \bigcap_{i}\pos{\aa _{i}} \mbox{ is feasible and }
% DAS
  \max_{\substack{\pp \in \CC \cap \bigcap_{i=1}^{n} \pos{\aa_{i}} \\
         \norm{\pp} = 1}}
% SHT
%  \max_{\pp \in \CC \cap \bigcap_{i=1}^{n} \pos{\aa_{i}} : \norm{\pp} = 1}
  \min_{i} \aa_{i}^{T} \pp < \epsilon
  }\\
\leq
  \prob{}{
   C \cap \bigcap_{i}\pos{\aa _{i}} \mbox{ is feasible and }
  \min_{i} \rho \left(\aa _{i}, \CC \cap \bigcap_{j \not = i} \pos{\aa_{j}} \right)
    < d \epsilon
  }.
\end{multline*}
Applying a union bound over $i$ and then Lemma~\ref{lem:wide-one},
  we find this probability is at most
\begin{align*}
\sum _{i=1}^{n}
\prob{}{
   C \cap \bigcap_{j}\pos{\aa _{j}} \mbox{ is feasible and }
  \rho \left(\aa _{i}, \CC \cap \bigcap_{j \not = i} \pos{\aa_{j}} \right)
    < d \epsilon
  }
& \leq \sum _{i=1}^{n} \frac{4 (\epsilon d) d^{1/4}}{\sigma }
 = \frac{4 n \epsilon d^{5/4}}{\sigma }.
\end{align*}

\end{proof}

%\begin{lemma}[Max of min of inner products is likely large]\label{lem:maxMinProb}
%Let $\CC$ be an open convex cone in $\Reals{d}$ and let
%  $\vs{\aa}{1}{n}$ be Gaussian random vectors of variance
%  $\sigma ^{2}$.
%Then,
%\[
%  \prob{}{
%   C \cap \bigcap_{i}\pos{\aa _{i}} \mbox{ is feasible and }
%  \max_{\substack{\pp \in \CC \cap \bigcap_{i=1}^{n} \pos{\aa_{i}} \\
%         \norm{\pp} = 1}}
%  \min_{i} \aa_{i}^{T} \pp < \epsilon
%  }
%  \leq
%  \left( \frac{4 \epsilon n d^{5/4}}{\sigma} \right).
%\]
%\end{lemma}
%\begin{proof}
%By Lemma~\ref{lem:maxMinBound},
%\begin{multline*}
%  \prob{}{
%   C \cap \bigcap_{i}\pos{\aa _{i}} \mbox{ is feasible and }
%  \max_{\substack{\pp \in \CC \cap \bigcap_{i=1}^{n} \pos{\aa_{i}} \\
%         \norm{\pp} = 1}}
%  \min_{i} \aa_{i}^{T} \pp < \epsilon
%  }\\
%\leq
%  \prob{}{
%   C \cap \bigcap_{i}\pos{\aa _{i}} \mbox{ is feasible and }
%  \min_{i} \rho \left(\aa _{i}, \CC \cap \bigcap_{j \not = i} \pos{\aa_{j}} \right)
%    < d \epsilon
%  }.
%\end{multline*}
%Applying first a union bound, and then Lemma~\ref{lem:wide-one},
%  we bound this by
%\begin{align*}
%\sum _{i=1}^{n}
%\prob{}{
%   C \cap \bigcap_{i}\pos{\aa _{i}} \mbox{ is feasible and }
%  \rho \left(\aa _{i}, \CC \cap \bigcap_{j \not = i} \pos{\aa_{j}} \right)
%    < d \epsilon
%  }
%& \leq \sum _{i=1}^{n} \frac{4 (\epsilon d) d^{1/4}}{\sigma }\\
%& = \frac{4 n \epsilon d^{5/4}}{\sigma }.
%\end{align*}

%\end{proof}

\begin{proof}[Proof of Lemma~\ref{lem:firstFeas}]
Follows immediately
  from Lemmas~\ref{lem:maxMin} and~\ref{lem:maxMinProb}.
\end{proof}
This concludes the analysis that it is unlikely that
  the primal program is both feasible and has small
  distance to ill-posedness.
Next, we show that it is unlikely that the primal
  program is both infeasible and has small distance to ill-posedness.

\subsection{Primal number, infeasible case}\label{ssec:primInfeas}

The main result of this subsection is:

\begin{lemma}[Infeasible is likely quite infeasible]\label{lem:firstInfeas}
Let $\CC$ be a non-pointed convex cone in $\Reals{d}$ and let
  $\orig{A}$ be any $n$-by-$d$ matrix such that
  $\fnorm{\orig{A}} \leq 1$.
Then, for any $0< \sigma \leq 1/\sqrt{d}$ and $\epsilon < 1/2$,
\begin{eqnarray*}
  \prob{A \leftarrow \calN (\orig{A},\sigma)}{
     \left( A \xx \geq 0, ~ \xx \in \CC
          \mbox{ is infeasible} \right)
     \mbox{ and }
     \left( \rho (A, \CC) \leq \epsilon  \right)}
%\lefteqn{  \prob{}{
%     \left( A \xx \geq 0, ~ \xx \in \CC
%          \mbox{ is infeasible} \right)
%     \mbox{ and }
%     \left( \rho (A, \CC) \leq \epsilon  \right)}} \\
& \leq &
    \frac{361~\epsilon n^{2} d^{1.5}
    {\log^{1.5} (1 /\epsilon )}}
%    \ceiling{\log^{1.5} (1 /\epsilon )}}
         {\sigma ^{2}}.
\end{eqnarray*}
\end{lemma}

To prove Lemma~\ref{lem:firstInfeas}, we consider adding the
  constraints one at a time.
If the program is infeasible in the end, then there must be some
  constraint, which we call the {\em critical} constraint,
  that takes it from being feasible to being infeasible.
Lemma~\ref{lem:firstInfeasGeom}
  gives a sufficient geometric
  condition for the program to be quite infeasible when the critical
  constraint is added.
We then prove Lemma~\ref{lem:firstInfeas}
  by showing that
  this condition is met with good probability.
The geometric condition is that the program is quite feasible
  before the critical constraint is added and that every previously
  feasible point is far from
  being feasible for the critical constraint.

\begin{lemma}[The feasible-to-infeasible transition]\label{lem:firstInfeasGeom}
Let $\CC$ be a non-pointed convex cone in $\Reals{d}$,
  $\pp \in \CC $ be a unit vector, and
  $\vs{\aa}{1}{k+1}$ be vectors in $\Reals{d}$ such that
\begin{align*}
  \form{\aa_{i}}{\pp}  \geq  \alpha,
     & \mbox{ for $1 \leq i \leq k$, and}\\
  \form{\aa_{k+1}}{\xx}  \leq  -\beta,
    &
      \mbox{ for all $\xx \in \CC
                      \intersect \bigcap_{i=1}^{k} \pos{\aa_{i}},
              ~\norm{\xx} = 1$}.
\end{align*}
Then,
\[
  \rho (\left[\vs{\aa}{1}{k+1}\right], \CC)
 \geq
  \min \setof{\frac{\alpha }{2},
              \frac{\alpha \beta }{4\alpha + 2\norm{\aa_{k+1}}}}.
\]
\end{lemma}

We will derive Lemma~\ref{lem:firstInfeasGeom} from
  the following geometric lemma.

\begin{lemma}[$\rho $ bound on inner product]\label{lem:infeasRhoIP}
Let $\CC $ be a non-pointed convex cone and let $\aa $ be a vector
  for which $\CC \cap \pos{\aa}$ is infeasible.
Then,
\[
  \max_{\pp \in \CC , \norm{\pp} = 1}
       \form{\pp }{\aa }
  \leq
    - \rho (\aa ,\CC ).
\]
\end{lemma}
\begin{proof}
Let $\pp$ be the unit vector in $\CC $ maximizing $\form{\pp }{\aa}$.
If we set
\[
  \Delta \aa = \left(\epsilon  - \form{\pp }{\aa} \right) \pp,
\]
for any $\epsilon > 0$, then
  we can see that $\CC \cap \pos{\aa + \Delta \aa }$
  is feasible from
\begin{align*}
  \form{\pp }{(\aa + \Delta \aa) }
& =
  \form{\pp }{\aa } +
  \left(\epsilon  - \form{\pp }{\aa} \right)
  \form{\pp }{ \pp}\\
& =
  \form{\pp }{\aa } +
  \left(\epsilon  - \form{\pp }{\aa} \right)\\
& =
\epsilon.
\end{align*}
So, we may conclude $\rho (\aa ,\CC ) \leq \abs{\form{\pp }{\aa}}$.
\end{proof}

\begin{proof}[Proof of Lemma~\ref{lem:firstInfeasGeom}]
The conditions of the lemma imply that
  $\CC \cap \bigcap _{i=1}^{k+1} \pos{\aa _{i}}$ is infeasible.
So, we may prove the lemma by demonstrating
  that for all $\epsilon $ satisfying
\begin{eqnarray}
  \epsilon & \leq & \alpha /2, \mbox{ and } \label{eqn:firstInfeasGeom1}\\
  \epsilon & < & \frac{\beta}{4 + 2\norm{\aa_{k+1}}/ \alpha},
     \label{eqn:firstInfeasGeom2}
\end{eqnarray}
and all $\setof{\vs{\Delta \aa}{1}{k+1}}$ satisfying
  $\norm{\Delta \aa_{i}} < \epsilon$ for $1 \leq i \leq k+1$, we have
\[
 \CC \intersect
   \bigcap_{i=1}^{k+1} \pos{\aa_{i} + \Delta \aa_{i}}
   \mbox{ is infeasible.}
\]

Assume by way of contradiction that
\[
  \CC \intersect
    \bigcap_{i=1}^{k+1} \pos{\aa_{i} + \Delta \aa_{i}}
    \mbox{ is feasible.}
\]
Then, there exists a unit vector
  $\xx ' \in \CC \intersect
   \bigcap_{i=1}^{k+1} \pos{\aa_{i} + \Delta \aa_{i}}$.
We first show that
\begin{equation}\label{eqn:firstInfeasGeom}
   \xx ' + \frac{\epsilon}{\alpha} \pp
  \in
   \CC \intersect \bigcap_{i=1}^{k} \pos{\aa_{i}}.
\end{equation}
To see this, consider any $i \leq k$ and note that
  $\form{(\aa_{i} + \Delta \aa_{i})}{\xx '} \geq 0$
  implies
\[
   \form{\aa_{i}}{\xx '} \geq -   \form{\Delta \aa_{i}}{\xx '}
  \geq
    - \norm{\Delta \aa_{i}} \norm{\xx '}
  \geq
    - \epsilon.
\]
Thus,
\[
  \form{\aa_{i}}{\left(\xx ' + \frac{\epsilon}{\alpha} \pp \right)}
  =
  \form{\aa_{i}}{\xx '} + \form{\aa_{i}}{ \frac{\epsilon}{\alpha} \pp}
  \geq
  - \epsilon + \frac{\epsilon}{\alpha} \alpha
  \geq
  0.
\]
To finish our proof of \eqref{eqn:firstInfeasGeom},
  we observe that $\xx' \in \CC$ and
  $\pp \in \CC$ imply
  $\xx' + \frac{\epsilon}{\alpha} \pp \in \CC$.

Let $\xx = \xx' + \frac{\epsilon}{\alpha} \pp$. Then
$\xx \in \CC \cap \bigcap_{i=1}^{k} \pos{\aa_{i}}$ and $\xx$ has
norm at most $1 + \epsilon / \alpha$  and at least
$1 - \epsilon / \alpha$.
To derive a contradiction, we now compute
\begin{eqnarray*}
  \form{(\aa_{k+1} + \Delta \aa_{k+1})}{\xx '}
& = &
  \form{(\aa_{k+1} + \Delta \aa_{k+1})}{(\xx - (\epsilon / \alpha) \pp)}\\
& = &
  \form{\aa_{k+1}}{\xx} + \form{\Delta \aa_{k+1}}{\xx }
  - (\epsilon / \alpha) \form{\aa_{k+1}}{\pp}
  - (\epsilon / \alpha) \form{\Delta \aa_{k+1}}{\pp}\\
& \leq &
  - \beta\norm{\xx }  + \norm{\Delta \aa_{k+1}}\norm{\xx }
  + (\epsilon / \alpha) \norm{\aa_{k+1}}
  + (\epsilon / \alpha ) \norm{\Delta \aa_{k+1}}\\
& \leq &
  - \beta (1 - \epsilon /\alpha ) + \epsilon (1 + \epsilon /\alpha )
  + (\epsilon / \alpha) \norm{\aa_{k+1}}
  + (\epsilon^{2} / \alpha ) \\
& = &
  - \beta (1 - \epsilon /\alpha ) +
   \epsilon \big(
      (1 + \epsilon / \alpha) + \norm{\aa_{k+1}}/ \alpha + \epsilon / \alpha
   \big)\\
& \leq  &
  - \beta/2 +
   \epsilon \left(
      2 + \norm{\aa_{k+1}}/ \alpha
   \right),
  \mbox{ by \eqref{eqn:firstInfeasGeom1}}\\
& <  &
  0
  \mbox{ by \eqref{eqn:firstInfeasGeom2}},
\end{eqnarray*}
which contradicts $\xx' \in \CC \intersect
   \bigcap_{i=1}^{k+1} \pos{\aa_{i} + \Delta \aa_{i}}$.
\end{proof}

We now prove that the geometric condition of
   Lemma \ref{lem:firstInfeasGeom} holds with high probability.
First, we establish two basic statements.

\begin{proposition}\label{pro:firstInfeasGeom}
For positive $\alpha $, $\beta $ and any vector $\aa _{k+1}$,
\[
   \frac{\alpha \beta }{2 \alpha + \norm{\aa _{k+1}}}
  \geq
   \min \setof{\frac{\alpha \beta }{2 + \norm{\aa _{k+1}}},
               \frac{\beta }{2 + \norm{\aa _{k+1}}}  }.
\]
\end{proposition}
\begin{proof}
For $\alpha \geq 1$, we have
\[
   \frac{\alpha \beta }{2 \alpha + \norm{\aa _{k+1}}}
  =
   \frac{\beta }{2 + \norm{\aa _{k+1}} / \alpha }
  \geq
   \frac{\beta }{2 + \norm{\aa _{k+1}} },
\]
while for $\alpha \leq 1$ we have
\[
   \frac{\alpha \beta }{2 \alpha + \norm{\aa _{k+1}}}
  \geq
   \frac{\alpha \beta }{2 + \norm{\aa _{k+1}}}.
\]
\end{proof}

\begin{proposition}\label{pro:infeasibleSubset}
If $\CC \intersect \bigcap _{i=1}^{k} \pos{\aa _{i}}$
  is infeasible, then
\[
  \rho \left([\vs{\aa }{1}{k}], \CC \right)
 \leq
  \rho \left([\vs{\aa }{1}{n}], \CC \right).
\]
\end{proposition}
\begin{proof}
Adding constraints cannot make it easier to change the program
  to make it feasible.
\end{proof}

\begin{proof}[Proof of Lemma~\ref{lem:firstInfeas}]
Let $\vs{\aa}{1}{n}$ be the rows of $A$, and let
\[
  \CC_{0} = \CC \quad \mbox{and} \quad \CC_{k} = \CC \intersect \bigcap_{i=1}^{k} \pos{\aa _{k}}.
\]
Note that $\CC_{n}$ is the final program.
Let $E_{k}$ denote the event that $\CC_{k-1}$ is feasible
  and $\CC_{k}$ is infeasible.
Using Proposition~\ref{pro:infeasibleSubset} and the fact that
  $\CC_{n}$ infeasible implies that $E_{k}$ must hold for some $k$,
  we obtain
\begin{align}
&  \prob{}{\mbox{$\CC_{n}$ is infeasible and
        $\rho \left([\vs{\aa}{1}{n}], \CC \right) \leq \epsilon $}}
  \nonumber \\
& \quad \leq  \sum_{k=0}^{n-1}
  \prob{}{\mbox{$E_{k+1}$ and
        $\rho \left([\vs{\aa}{1}{n}], \CC \right) \leq \epsilon $}}
 \nonumber \\
& \quad \leq  \sum_{k=0}^{n-1}
  \prob{}{\mbox{$E_{k+1}$ and
        $\rho \left([\vs{\aa}{1}{k+1}], \CC \right) \leq \epsilon$}}.
  \label{eqn:firstInfeas0}
\end{align}
If $E_{k+1}$ occurs, then $\CC_{k}$ is feasible, and we may define
\[
  \kappa (\vs{\aa }{1}{k})
 =
  \max_{\substack{\pp \in \CC _{k} \\
     \norm{\pp} = 1}}
   \min _{1 \leq i \leq k} \form{\aa_{i}}{\pp}.
\]
Then, $E_{k+1}$ implies
\begin{align*}
  \form{\aa _{i}}{\pp }
  & \geq
  \kappa (\vs{\aa }{1}{k}),
  \mbox{ for $1 \leq i \leq k$, and}
\\
\intertext{Lemma~\ref{lem:infeasRhoIP} implies}
  \form{\aa _{k+1}}{\xx }
  & \leq
  - \rho (\aa _{k+1}, \CC _{k})
  \mbox{ for all $\xx \in \CC _{k}$, $\norm{\xx } = 1$}.
\end{align*}
So, we may apply
  Lemma~\ref{lem:firstInfeasGeom} and
  Proposition~\ref{pro:firstInfeasGeom}
  to show that
  $E_{k+1}$ implies
\begin{align}
  \rho \left([\vs{\aa}{1}{k+1}], \CC \right)
  & \geq
  \min \setof{
    \frac{\kappa (\vs{\aa}{1}{k}) }{2},
    \frac{\kappa (\vs{\aa}{1}{k}) \rho (\aa_{k+1}, \CC_{k}) }
         {4 + 2\norm{\aa _{k+1}}},
    \frac{\rho (\aa_{k+1}, \CC_{k}) }{4 + 2\norm{\aa _{k+1}}}
   }\nonumber \\
& \geq
\frac{
  \min \setof{
    \kappa (\vs{\aa}{1}{k}) ,~
    \kappa (\vs{\aa}{1}{k}) \rho (\aa_{k+1}, \CC_{k}),~
    \rho (\aa_{k+1}, \CC_{k})
   }
}{
   4 + 2\norm{\aa _{k+1}}
} \label{eqn:firstInfeas1}
\end{align}

We now proceed to bound the probability that
  the numerator of this fraction is small.

We first note that
\[
  \kappa (\vs{\aa}{1}{k}) \rho (\aa_{k+1}, \CC_{k})  \leq \lambda
\]
implies that either
  $\kappa (\vs{\aa}{1}{k}) \leq \lambda $,
  $\rho (\aa_{k+1}, \CC_{k}) \leq \lambda $, or there exists
  an $l$ between 1 and $\ceiling{\log (1/\lambda )}$ for which
\[
  \kappa (\vs{\aa}{1}{k}) \leq 2^{-l+1} \mbox{ and }
  \rho (\aa_{k+1}, \CC_{k}) \leq 2^{l} \lambda.
\]
We apply Lemma~\ref{lem:wide-one} to bound
\begin{equation}\label{eqn:primalfoo1}
   \prob{\aa_{k+1}}{E_{k+1} \mbox{ and } \rho (\aa_{k+1}, \CC_{k}) \leq \lambda}
  \leq
    \frac{4 \lambda d^{1/4}}{\sigma },
\end{equation}
and Lemma~\ref{lem:maxMinProb} to bound
\begin{equation}\label{eqn:primalfoo2}
  \prob{\vs{\aa}{1}{k}}
       {\mbox{$\CC_{k}$ is feasible and
        $\kappa (\vs{\aa}{1}{k}) \leq \lambda$}}
 \leq
  \frac{4 \lambda n d^{5/4}}{\sigma }.
\end{equation}

So, for $1 \leq l \leq \ceiling{\log (1/\lambda )}$, we obtain
\begin{eqnarray*}
\lefteqn{\prob{\vs{\aa}{1}{k+1}}
     {\mbox{$E_{k+1}$  and }
  \kappa (\vs{\aa}{1}{k}) \leq 2^{-l+1}
   \mbox{ and }
  \rho (\aa_{k+1}, \CC_{k}) \leq 2^{l} \lambda}}\\
& = &
\prob{\vs{\aa}{1}{k}}
     {\mbox{$\CC_{k} \not = \emptyset$  and }
      \kappa (\vs{\aa}{1}{k}) \leq 2^{-l+1}}  \cdot \\
& &
  \prob{\aa _{k+1}}
     {\mbox{ $\CC_{k+1} = \emptyset$  and }
     \rho (\aa_{k+1}, \CC_{k}) \leq 2^{l} \lambda
    ~~|~~ \mbox{$\CC_{k} \not = \emptyset$  and }
      \kappa (\vs{\aa}{1}{k}) \leq 2^{-l+1}}
\\
& \leq &
\prob{\vs{\aa}{1}{k}}
     {\mbox{$\CC_{k} \not = \emptyset$  and }
      \kappa (\vs{\aa}{1}{k}) \leq 2^{-l+1}}
   \frac{2^{l} 4 \lambda d^{1/4}}{\sigma }~,
  \mbox{ by \eqref{eqn:primalfoo1}}
\\
& \leq &
  \frac{2^{-l+1} 4 n d^{5/4}}{\sigma }
   \frac{2^{l} 4 \lambda d^{1/4}}{\sigma }~,
  \mbox{ by \eqref{eqn:primalfoo2}, }
\\
& = &
  \frac{32 \lambda n d^{1.5}}
       {\sigma ^{2}}.
\end{eqnarray*}
Summing over the choices for $l$, we obtain
\begin{eqnarray}
\lefteqn{  \prob{}
       {E_{k+1} \mbox{ and }
    \min \setof{
      \kappa (\vs{\aa}{1}{k}) ,
      \kappa (\vs{\aa}{1}{k}) \rho (\aa_{k+1}, \CC_{k}),
      \rho (\aa_{k+1}, \CC_{k})
     } < \lambda
  }
} \nonumber \hspace{1in}\\
&  \leq &
     \frac{4 \lambda n d^{5/4}}{\sigma } +
     \frac{4 \lambda d^{1/4}}{\sigma } +
   \ceiling{\log (1/\lambda ) }
     \frac{32 \lambda n d^{1.5}}
          {\sigma ^{2}} \nonumber \\
& \leq &
  \lambda \left(
    \frac{4 n d^{3/4} + 4 +
            32 \ceiling{\log (1/\lambda )} n d^{1.5}}
         {\sigma ^{2}}
   \right),
  \mbox{ by $\sigma \leq 1/\sqrt{d},$} \nonumber \\
& \leq &
  \lambda \left(
    \frac{(32 \ceiling{\log (1/\lambda )} + 8)  n d^{1.5}}
         {\sigma ^{2}}
   \right). \label{eqn:firstInfeas2}
\end{eqnarray}

This concludes our analysis of the numerator of (\ref{eqn:firstInfeas1}).
We can bound the probability that the
  denominator of \eqref{eqn:firstInfeas1} is small by observing
  that $\aa _{k+1}$ is a Gaussian centered at a point
  $\orig{\aa} _{k+1}$ of norm at most 1; so,
  Corollary~\ref{cor:chiSquare} implies
\begin{equation}\label{eqn:firstInfeas3}
  \prob{}{4 + 2\norm{\aa _{k+1}} \geq
          6 + 2\sigma \sqrt{2d  \ln (e/ \epsilon  )}} \leq  \epsilon .
\end{equation}
We now set
  $\lambda = \epsilon (6 + 2 \sigma \sqrt{2 d \ln (e/ \epsilon  )})$
  and observe that if we had
\[
\frac{
  \min \setof{
    \kappa (\vs{\aa}{1}{k}) ,~
    \kappa (\vs{\aa}{1}{k}) \rho (\aa_{k+1}, \CC_{k}),~
    \rho (\aa_{k+1}, \CC_{k})
   }
}{
   4 + 2\norm{\aa _{k+1}}
} \leq \epsilon
\]
this would imply
\begin{align*}
  \min \setof{
    \kappa (\vs{\aa}{1}{k}) ,~
    \kappa (\vs{\aa}{1}{k}) \rho (\aa_{k+1}, \CC_{k}),~
    \rho (\aa_{k+1}, \CC_{k})} & < \lambda,
\mbox{ or }\\
 4 + 2\norm{\aa _{k+1}} & \geq
          6 + 2\sigma \sqrt{2d  \ln (e/ \epsilon  )}.
\end{align*}
So, we may apply \eqref{eqn:firstInfeas2}
  and \eqref{eqn:firstInfeas3} to obtain
\begin{align*}
\lefteqn{  \prob{}{E_{k+1} \mbox{ and }
          \rho \left([\vs{\aa}{1}{k+1}], \CC \right) \leq \epsilon}}
\\
&  \leq
   \epsilon +
   \epsilon
       \left(6 + 2 \sigma \sqrt{2 d  \ln (e/ \epsilon  )} \right)
   \left(
    \frac{(32
              \ceiling{\log (1 / \epsilon
                             (6 + 2\sigma \sqrt{d \log (e/ \epsilon)}) )}
              +8)n d^{1.5}}
         {\sigma ^{2}}
   \right) \\
& \leq
   \epsilon +
   \epsilon
       \left(6 + 3 \sqrt{\ln (e/ \epsilon  )} \right)
   \left(
    \frac{(32
              \ceiling{\log (1 /(6\epsilon) )}
             +8) n d^{1.5}}
         {\sigma ^{2}}
   \right), \text{ using $\sigma \leq 1/\sqrt{d}$ in the first term} \\
& \leq
   \epsilon +
       \epsilon \left(9  \sqrt{\ln (e/ \epsilon  )} \right)
   \left(
    \frac{(32
              \ceiling{\log (1 /(6\epsilon) )}
              +8)n d^{1.5}}
         {\sigma ^{2}}
   \right) \\
& \leq
   \epsilon +
       \epsilon
   \left(
    \frac{360
              {\log^{1.5} (1 /\epsilon )}
              n d^{1.5}}
         {\sigma ^{2}}
   \right),  \\
\intertext{since $(\sqrt{\ln (e/\epsilon)}) (\ceiling{\log{(1/6\epsilon)}}+1/4) \leq
          \log^{1.5} (1/\epsilon)$ for $\epsilon < 1/361$,}
& \leq
  \epsilon
   \left(
    \frac{361
              {\log^{1.5} (1 /\epsilon )}
              n d^{1.5}}
         {\sigma ^{2}}
   \right) 
\end{align*}
Plugging this in to (\ref{eqn:firstInfeas0}), we get
\begin{eqnarray*}
  \prob{}{\mbox{$\CC_{0}$ is infeasible and
        $\rho \left([\vs{\aa}{1}{n}], \CC \right) \leq \epsilon $}}
& \leq &
    \frac{ 361 \epsilon n^{2} d^{1.5}
              {\log^{1.5} (1 /\epsilon )}}
         {\sigma ^{2}}.
\end{eqnarray*}
\end{proof}

\subsection{Primal condition number, putting the feasible and infeasible cases together}\label{ssec:primBoth}

We combine the results of
  Sections~\ref{ssec:primFeas} and~\ref{ssec:primInfeas}
  to prove Lemma~\ref{lem:primal}, which says that the primal condition number
  is probably low.

\begin{proof}[Proof of Lemma~\ref{lem:primal}]
In Lemma~\ref{lem:firstFeas}, we show that
\[
  \prob{}{
     \left( A \xx \geq 0, ~ \xx \in \CC
          \mbox{ is feasible} \right)
     \mbox{ and }
     \left( \rho (A, \CC) \leq \epsilon \right)}
  \leq
   \frac{4 \epsilon n d^{5/4}}{\sigma},
\]
while in Lemma~\ref{lem:firstInfeas}, we show
\begin{eqnarray*}
\prob{}{
     \left( A \xx \geq 0, ~ \xx \in \CC
          \mbox{ is infeasible} \right)
     \mbox{ and }
     \left( \rho (A, \CC) \leq \epsilon  \right)}
& \leq &
    \frac{361 \epsilon {\log^{1.5} (1 /\epsilon )} n^{2} d^{1.5}}
         {\sigma ^{2}}.
\end{eqnarray*}
Thus,
\begin{eqnarray}
\prob{}{\rho (A, \CC ) \leq \epsilon }
& = &
  \prob{}{
     \left( A \xx \geq 0, ~ \xx \in \CC
          \mbox{ is feasible} \right)
     \mbox{ and }
     \left( \rho (A, \CC) \leq \epsilon  \right)} \nonumber \\
& &
 +~~
 \prob{}{
     \left( A \xx \geq 0, ~ \xx \in \CC
          \mbox{ is infeasible} \right)
     \mbox{ and }
     \left( \rho (A, \CC) \leq \epsilon   \right)} \nonumber\\
& \leq  &
   \frac{4 \epsilon n d^{5/4}}{\sigma}
    +
    \frac{361 \epsilon {\log^{1.5} (1 /\epsilon )} n^{2} d^{1.5}}
         {\sigma ^{2}}
   \nonumber \\
& \leq &
    \frac{365 \epsilon {\log^{1.5} (1 /\epsilon )} n^{2} d^{1.5}}
         {\sigma ^{2}}
   \label{eqn:boundonrho}
\end{eqnarray}
Setting $\epsilon = \delta/(3 \alpha \log^{1.5} (\alpha/ \delta))$ where
  $\alpha = 365 \frac{n^{2} d^{1.5}}{\sigma ^{2}}$ (note that this satisfies $\epsilon < 1/2$),
  we obtain
\begin{eqnarray}
   \prob{}{\frac{1}{\rho(A, \CC)} \geq
\frac{1100~n^{2} d^{1.5} } { \delta \sigma^2}
\log^{1.5} \left(\frac{365~n^2 d^{1.5}}{\delta \sigma^2}\right)
} & \leq & \frac{\alpha \delta \log^{1.5}
 \left(\frac{3\alpha}{\delta} \log^{1.5} \left(\frac{\alpha}{\delta}\right)\right)
        }{3\alpha \log^{1.5} (\frac{\alpha}{\delta})} \nonumber
\\ & \leq & 0.74 ~ \delta ,    \label{eqn:primal1}
\end{eqnarray}
as $\alpha / \delta \geq 365$.

At the same time, Corollary~\ref{cor:chiSquare} tells us that
\[
   \prob{}{\fnorm{A} \geq 1 + \sigma \sqrt{nd~ 2 \ln (4e/\delta )}}
 \leq
   \delta/4.
\]
The lemma now follows by applying this bound, $\sigma \leq 1/\sqrt{nd}$,
  and (\ref{eqn:primal1}), to get
\begin{eqnarray*}
   \prob{}{\frac{\fnorm{A}}{\rho(A, \CC)} \geq
\frac{
(1 + \sqrt{2 \ln (4e/\delta )})
1100~ n^{2} d^{1.5} } { \delta \sigma^2}
\log^{1.5} \left(\frac{365~ n^2 d^{1.5}}{\delta \sigma^2}\right)}
\leq (0.74 + 0.25) \delta < \delta
\end{eqnarray*}
To derive the lemma as stated, we note
\[
\frac{
(1 + \sqrt{2 \ln (4e/\delta )})
1100~ n^{2} d^{1.5} } { \delta \sigma^2}
\log^{1.5} \left(\frac{365~ n^2 d^{1.5}}{\delta \sigma^2}\right)
\leq
  \frac{2^{12}~n^{2} d^{1.5}}{\delta \sigma^2}
  \log^2 \left(\frac{2^{9}~ n^2 d^{1.5}}{\delta \sigma^2}\right).
\]
\end{proof}

\subsection{Log of the Primal Condition Number}\label{ssec:logexp}

In this section, we prove we prove Lemma \ref{lem:logcond}.

\begin{proof}[Proof of Lemma~\ref{lem:logcond}]
First notice that
\begin{eqnarray*}
\expec{}
     {\log \frac{\fnorm{A}}{\rho (A, \CC )}}
 = \expec{}
     {\log \fnorm{A} +  \log \frac{1}{\rho (A, \CC )}}.
\end{eqnarray*}

We first focus on $\expec{}
     {\log \fnorm{A}}$.
Because logarithm is a convex function, we have
\[
\expec{}{\log \fnorm{A}} \leq
 \log ( \expec{}{\fnorm{A}})
\leq 
  \log \sqrt{
  \expec{}{\fnorm{A}^{2}}}.
\]
As $\fnorm{A}^{2}$ is a
  $dn$-dimensional
  non-central $\chi^{2}$ random variable
  with non-centrality parameter  $\fnorm{\orig{A}}$, its expectation
  is $nd + \fnorm{\orig{A}}$~\cite[26.4.37]{AbramowitzStegun}.
Therefore,
\[
\expec{}{\log \fnorm{A}} \leq
 \log \sqrt{nd+1}.
\]

%{\bf \} }

We will use the following simple fact which is easy to verify numerically:

\begin{fact}\label{fac:spielmantrick}
For all $\alpha \geq 100$ and $x \geq
2\log \alpha $, $x -1.5\log x \geq x/2$.
\end{fact}

Let
\[
\alpha = \frac{365n^{2}d^{1.5}}{\sigma^{2}},
\]
as before. By Equation (\ref{eqn:boundonrho}) in the proof of
   Lemma \ref{lem:primal},
\[
\prob{}{\frac{1}{\rho (A, \CC )} \geq x} \leq \frac{\alpha \log^{1.5}x}{x}.
\]
Therefore,
\begin{eqnarray*}
\expec{}{\log \frac{1}{\rho (A, \CC )}} & = & \int_{0}^{\infty}
\prob{}{\log \frac{1}{\rho (A, \CC )} > x} dx  \\
& =&  \int_{0}^{\infty}
\prob{}{\frac{1}{\rho (A, \CC )} > e^{x}} dx \\
& \leq & \int_{0}^{\infty}
 \min \left (1,\frac{\alpha x^{1.5}}{e^{x}}\right)dx\\
& \leq & \int_{0}^{2\log \alpha }dx + \int_{2\log \alpha }^{\infty}
  \frac{\alpha x^{1.5}}{e^{x}} dx\\
& = & 2\log \alpha + \alpha \int_{2\log \alpha }^{\infty}
  e^{-x +1.5\log x} dx \\
& \leq & 2\log \alpha + \alpha \int_{2\log \alpha }^{\infty}
e^{-x/2}dx \\
& \leq  & 2\log \alpha + 2,
\end{eqnarray*}
where the second-to-last inequality follows from Fact \ref{fac:spielmantrick}.

Thus,
\begin{eqnarray*}
\expec{}
     {\log \frac{\fnorm{A}}{\rho (A, \CC )}}
& = & \expec{}
     {\log \fnorm{A} +  \log \frac{1}{\rho (A, \CC )}} \leq \log
     \sqrt{nd+1} + 2\log \alpha + 2 \\
& \leq  & 14 + 4.5 \log \frac{nd}{\sigma }.
\end{eqnarray*}
\end{proof}

% Local Variables: ***
% TeX-master:"journalipm.tex" ***
% End: ***

\section{Dual Condition Number}\label{sec:dual}

\setcounter{theorem}{0}

In this section, we consider linear programs
of the form
\[
  A^{T} \yy  = \cc , ~ \yy  \geq \bvec{0}.
\]

The dual program of form~(1) and the primal program of form~(3) are both
of this type. The dual program of form~(4) can be handled using a
slightly different argument than the one we present. As in section
\ref{sec:primal}, we omit the details of the modifications necessary for
form~(4). We begin by defining distance to
ill-posedness appropriately for the form of linear program
considered in this section:

\begin{definition}[Dual distance to ill-posedness]\label{def:rho3}
For a matrix, $A$, and a vector $\cc$, we define $\rho (A, \cc )$ by
\begin{itemize}
\item [a.]
  if $A^{T} \yy  = \cc , ~ \yy  \geq \bvec{0}$ is feasible, then $\rho(A , \cc )
 = $
\begin{eqnarray*}
%\hspace{-.3in}
 \sup \setof{\epsilon : \fnorm{\Delta A} + \fnorm{\Delta \cc}
    < \epsilon \mbox{ implies }
          (A + \Delta A)^{T} \yy = \cc + \Delta \cc, ~ \yy  \geq \bvec{0}
    \mbox{ is feasible}}
\end{eqnarray*}
\item [b.]
  if $A^{T} \yy  = \cc , ~ \yy  \geq \bvec{0}$ is infeasible, then
$\rho(A , \cc )   = $
\begin{eqnarray*}
%\hspace{-.3in}
 \sup \setof{\epsilon : \fnorm{\Delta A} + \fnorm{\Delta \cc}
    < \epsilon \mbox{ implies }
          (A + \Delta A)^{T} \yy = \cc + \Delta \cc, ~ \yy  \geq \bvec{0}
    \mbox{ is infeasible}}
\end{eqnarray*}
\end{itemize}
\end{definition}

The main result of this section is:
\begin{lemma}[Dual condition number is likely low]\label{lem:dual2}
Let $\orig{A}$ be an $n$-by-$d$ matrix and $\cc $ be a vector in
$\Reals{d}$
  such that $\fnorm{\orig{A}} \leq 1$ and $\norm{\orig{\cc }} \leq 1$.
Then for any $\sigma \leq 1/\sqrt{nd}$,
\begin{eqnarray*}
\prob{(A,\cc )
  \leftarrow \calN ((\orig{A},\orig{\cc }),\sigma )}
  {\frac{\fnorm{A, \cc}}{\rho(A, \cc )}
        >
    \frac{50000~ d^{1/4} n^{1/2}}{\epsilon \sigma^2}
    \log^{2} \left( \frac{200~d^{1/4} n^{1/2}}{\epsilon \sigma^2} \right)
    }
& \leq & \epsilon.
\end{eqnarray*}
In addition,
\begin{eqnarray*}
\expec{(A,\cc )
  \leftarrow \calN ((\orig{A},\orig{\cc }),\sigma )}
   {\log \frac{\fnorm{A, \cc}}{\rho(A, \cc )}}
 \leq 14 + 4\log \frac{nd}{\sigma}.
\end{eqnarray*}

\end{lemma}

We begin by giving several common definitions that will be useful
in our analysis of the dual condition number
(Section~\ref{ssec:basic}). We define a change of variables
(Section~\ref{ssec:change}), and we then develop a sufficient
geometric condition for the dual condition number to be low
(Section~\ref{ssec:dualcond}). In Section~\ref{ssec:dual-prob}, we
use Lemma \ref{lem:dualProb} to prove Lemma \ref{lem:dual2}, thereby
establishing that this geometric condition is met with good probability.

\subsection{Geometric Basics}\label{ssec:basic}

\begin{definition}[Cone]\label{def:cone2}
For a set of vectors $\vp{\aa}{1}{n}$, let $\cone{\vp{\aa}{1}{n}}$
  denote\\
  $\setof{\xx : \xx = \sum_i \lambda_i \aa_i, ~~\lambda_i \geq 0 }.$
\end{definition}

\begin{definition}[Hull]\label{def:hull}
For a set of vectors $\vp{\aa}{1}{n}$, let $\hull{\vp{\aa}{1}{n}}$
  denote\\
  $\setof{\xx : \xx = \sum_i \lambda_i \aa_i, ~~\lambda_i \geq 0, ~~\sum_i
  \lambda_i = 1}.$
\end{definition}

\begin{definition}[Boundary of a set]\label{def:bdry}
For a convex set $\SS$, let $\bdry{\SS}$ denote the boundary of $\SS$,
  i.e.,
  $\setof{\xx : \forall \epsilon > 0, ~\exists \ee, ~\norm{\ee} \leq \epsilon,
  ~s.t.~ \xx + \ee \in \SS, ~\xx - \ee \notin \SS}.$
\end{definition}

\begin{definition}[Point-to-set distance]\label{def:distance}
Let $\dist{\xx}{\SS}$ denote the distance of $\xx$ to $\SS$, i.e.,\\
  $\min \setof{\epsilon   : \exists \ee, ~\norm{\ee} \leq \epsilon,
  ~s.t.~ \xx + \ee \in \SS}.$
\end{definition}

Note that $\cone{\vp{\aa}{1}{n}}$ is \emph{not} a non-pointed convex cone,
while $\hull{\vp{\aa}{1}{n}}$ is the standard convex hull of
$\{\vp{\aa}{1}{n}\}$.

\subsection{Change of variables}\label{ssec:change}

We observe that there exists a solution to
 the system $A^{T} \yy  = \cc , ~ \yy  \geq \bvec{0}$ if and only if
\[
  \cc \in \cone{\vp{\aa}{1}{n}},
\]
and that for $\cc \neq \bvec{0},$ this holds if and only if
\[
  \ray{\cc } \intersects \hull{\vp{\aa}{1}{n}}.
\]

In this Section, we need one technique beyond those used in 
  Section~\ref{sec:primal}---a change of variables.
We set
\begin{eqnarray*}
 \zz & = & (1/n) \sum_{i=1}^{n} \aa _{i} , \mbox{ and }\\
 \xx _{i} & = & \aa _{i} - \zz , \mbox{ for $i = 1$ to $n-1$}.
\end{eqnarray*}
For notational convenience, we let
  $\xx _{n} = \aa _{n} - \zz $, although $\xx_n$ is not independent of
  $\setof{\zz, \vp{\xx}{1}{n-1}}$.

We can restate the condition for the linear program to be
  ill-posed in these new variables:
\begin{lemma}[Ill-posedness in new variables]\label{lem:cov}
\[
  A^{T} \yy  = \cc , ~ \yy  \geq \bvec{0}, ~ \cc \neq \bvec{0}
  \mbox{ is ill-posed if and only if }
  \zz \in \bdry{\ray{\cc} -  \hull{\vp{\xx}{1}{n}}}.
\]
\end{lemma}
\begin{proof}
We observe
\begin{eqnarray*}
  A^{T} \yy  = \cc , ~ \yy  \geq \bvec{0} \mbox{ is feasible }
 & \iff &
  \ray{\cc } \intersects \hull{\vp{\aa}{1}{n}}\\
 & \iff &
  \ray{\cc } \intersects \zz + \hull{\vp{\xx}{1}{n}}\\
 & \iff &
  \zz \in \ray{\cc} -  \hull{\vp{\xx}{1}{n}}.
\end{eqnarray*}
For $\cc \neq \bvec{0}$, $\ray{\cc} -  \hull{\vp{\xx}{1}{n}}$ is a
continuous mapping from $\cc, \vp{\xx}{1}{n}$ to subsets of Euclidean
space, and so for $\zz$ in the set and not on the boundary, a
sufficiently small
change to all the variables simultaneously will always leave
$\zz$ in the set, and similarly for $\zz$ not in the set and not on the
boundary.

To establish the other direction, we observe that
  if $\zz$ is on the boundary, then
  can perturb $\zz$ to bring it in or out of the set. Although
$\zz, \vp{\xx}{1}{n}$ are determined by the $\vp{\aa}{1}{n}$, we
can perturb the $\vp{\aa}{1}{n}$ so as to change the value of
$\zz$ without changing the values of any of the  $\vp{\xx}{1}{n}$.
This can be done because each $\xx_i$ is a relative offset from
the average $\zz$, while each $\aa_i$ is an absolute offset from
the origin; the proof of lemma~\ref{lem:zGaussian} below
establishes formally that the change of variables permits this.

The lemma is also true for $\cc = \bvec{0}$, but we will not need this
  fact.
\end{proof}

Note that $\ray{\cc} -  \hull{\vp{\xx}{1}{n}}$ is a convex set.
The following lemma will allow us to apply lemma~\ref{lem:small}
  to determine the probability that $\zz$ is near the
  boundary of this convex set.

\begin{lemma}[Independence of mean among new
  variables]\label{lem:zGaussian}
Let $\vp{\orig{\aa}}{1}{n}$ be $n$ vectors in $\Reals{d}$.
Let $\vp{\aa}{1}{n}$ be a Gaussian perturbation
   of $\vp{\orig{\aa}}{1}{n}$ of variance $\sigma ^{2}$.
Let
\[
  \zz = \frac{1}{n}\sum _{i}\aa _{i}
  \mbox{ and }
  \xx _{i} = \aa _{i} - \zz ,
  \mbox{ for $1 \leq i \leq n$}.
\]
Then, $\zz $ is a Gaussian perturbation of
\[
\orig{\zz } = \frac{1}{n}\sum _{i}\orig{\aa} _{i},
\]
of variance $\sigma ^{2} / n$ and is independent of $\vp{\xx }{1}{n}$.
\end{lemma}
\begin{proof}
As $\zz $ is the average of Gaussian perturbations of variance
   $\sigma^{2}$ of $n$ vectors $\vp{\orig{\aa}}{1}{n}$,
  it is a Gaussian perturbation of variance
  $\sigma ^{2}/n$ of the average of these $n$ vectors, that is, of
\[
\orig{\zz } = \frac{1}{n}\sum _{i}\orig{\aa} _{i}.
\]
The vector $\zz $ is independent of $\vp{\xx}{1}{n}$
  because the linear combination of $\vp{\aa}{1}{n}$
  used to obtain $\zz$ is orthogonal
  to the linear combinations of $\vp{\aa}{1}{n}$
  used to obtain the $\xx _{i}$s.
\end{proof}

\begin{lemma}[Mean is likely far from
  ill-posedness]\label{lem:dualProb}
Let $\vp{\orig{\aa}}{1}{n}$ be $n$ vectors in $\Reals{d}$ and
  $\orig{\cc} $ be a vector in $\Reals{d}$.
Let $\vp{\aa}{1}{n}$ be a Gaussian perturbation
  of $\vp{\orig{\aa}}{1}{n}$ of variance $\sigma ^{2}$ and let $\cc $
  be a Gaussian perturbation of $\orig{\cc }$ of variance $\sigma^{2}$.
Let
\[
  \zz = \frac{1}{n}\sum _{i}\aa _{i}
  \mbox{ and }
  \xx _{i} = \aa _{i} - \zz ,
  \mbox{ for $1 \leq i \leq n$}.
\]
Then, for all $\cc$ and $\vs{\xx}{1}{n}$,
\[
  \prob{\zz}{\dist{\zz}{\bdry{\ray{\cc} - \hull{\vp{\xx}{1}{n}}}} \leq \epsilon}
 \leq
   \frac{8 \epsilon d^{1/4}n^{1/2}}{\sigma} .
\]
\end{lemma}
\begin{proof}
Let $\cc$ be arbitrary.
By Lemma~\ref{lem:zGaussian}, we can choose $\vp{\xx }{1}{n}$
  and then choose $\zz $ independently.
Having chosen $\vp{\xx}{1}{n}$, we fix the convex body
  $\ray{\cc} - \hull{\vp{\xx}{1}{n}}$
  and apply Lemma~\ref{lem:small} twice:
  once for the inside $\epsilon$-boundary, and once
  for the outside $\epsilon$-boundary.
\end{proof}

\subsection{A geometric characterization of dual condition number}
  \label{ssec:dualcond}

We now give a geometric characterization of the dual condition number
   that uses both the original and the new variables.
In the next section, we will
  use this characterization
  to  prove Lemma~\ref{lem:dual2}.

\begin{lemma}[Reciprocal of distance to ill-posedness]\label{lem:dualGeom}
Let $\cc $ and $\vp{\aa}{1}{n}$ be vectors in $\Reals{d}$.
Let
\[
  \zz = \frac{1}{n}\sum _{i}\aa _{i}
  \mbox{ and }
  \xx _{i} = \aa _{i} - \zz ,
  \mbox{ for $1 \leq i \leq n$}.
\]
\[
 k_1 = \dist{\zz}{\bdry{\ray{\cc} - \hull{\vp{\xx}{1}{n}}}}
\]
\[
 k_2 = \norm{\cc }
\]
Then
\[
  \frac{1}{\rho(A, \cc)} \leq
    \max\setof{\frac{8}{k_1}, \frac{4}{k_2}, \frac{24 \max_i{\norm{\aa_i}}}{k_1 k_2}}.
\]
\end{lemma}
\begin{proof}
By the definition of $k_1$ and $k_2$ and Lemma \ref{lem:secondGeom}, we
can tolerate any change of magnitude up to $k_1/4$ in $\zz$, and
  $\vs{\xx}{1}{n}$, and any
change of up to $\frac{k_1 k_2}{2k_1 + 4(\norm{\zz} + \max
\norm{\xx_i})}$ in $\cc$ without the program becoming ill-posed. We show
that this means we can tolerate any change of up to $k_1/8$ in $\aa_i$
without the program becoming ill-posed. Formally, we need to show that
if $\norm{\Delta \aa_i} \leq k_1/8$ for all $i$, then $\norm{\Delta \zz} \leq
k_1/4$ and $\norm{\Delta \xx_i} \leq k_1/4$. Since $\Delta \zz = (1/n)
\sum \Delta \aa_i$, $\norm{\Delta \zz} \leq k_1/8$. Since $\Delta \xx_i
= \Delta \aa_i - \Delta \zz$, $\norm{\Delta \xx_i} \leq k_1/8 + k_1/8
= k_1/4$. Thus
$$ \rho(A,\cc) \geq \min\setof{\frac{k_1}{8} ,
\frac{k_1 k_2}{2k_1 + 4(\norm{\zz} + \max
\norm{\xx_i})}
} $$
which implies
$$ \frac{1}{\rho(A,\cc)} \leq \max\setof{\frac{8}{k_1} ,
\frac{4}{k_2},
\frac{8(\norm{\zz} + \max \norm{\xx_i})}{k_1 k_2}
}, $$
as
\[
\frac{2k_1 + 4(\norm{\zz} + \max \norm{\xx_i})}
     {k_1 k_2}
\leq
\begin{cases}
\frac{4}{k_{2}}
& \mbox{ if $k_{1} \geq 4(\norm{\zz} + \max \norm{\xx_i})$, and}\\
\frac{8(\norm{\zz} + \max \norm{\xx_i})}{k_1 k_2}
& \mbox{otherwise}.
\end{cases}
\]
Since 
 $\zz = (1/n)\sum \aa_i $ implies $\norm{\zz} \leq
 \max\norm{\aa_i}$, and $\xx_i = \aa_i - \zz $  implies $
\norm{\xx_i} \leq  \norm{\aa_i} + \norm{\zz} \leq 2 \max\norm{\aa_i}$, we
have
$$ \frac{1}{\rho(A,\cc)} \leq \max\setof{\frac{8}{k_1} ,
\frac{4}{k_2},
\frac{24\max \norm{\aa_i}}{k_1 k_2}
}. $$
\end{proof}

\begin{lemma}\label{lem:secondGeom}{\bf (Geometric condition to be far
from ill-posedness in new variables.)}
If
\begin{eqnarray}
  \dist{\zz}{\bdry{\ray{\cc} - \hull{\vp{\xx}{1}{n}}}} > \alpha \label{foo:3}
\end{eqnarray}
and
\begin{eqnarray*}
  \norm{\Delta \xx_{i}} & \leq & \alpha /4, \\
  \norm{\Delta \zz} & \leq & \alpha /4,\\
  \norm{\Delta \cc } & \leq &
   \frac{\alpha \norm{\cc}}
        {2 \alpha + 4 (\norm{\zz} + \max_{i} \norm{\xx_{i}})}, \\
\end{eqnarray*}
then
\[
\zz  + \Delta \zz  \not \in
  \bdry{\ray{\cc + \Delta \cc } -
  \hull{\xx_{1} + \Delta \xx_{1}, \ldots , \xx_{n} + \Delta \xx_{n}}}
\]
\end{lemma}

\begin{proof}
Assume by way of contradiction that
\[
\zz  + \Delta \zz  \in
  \bdry{\ray{\cc + \Delta \cc } -
  \hull{\xx_{1} + \Delta \xx_{1}, \ldots , \xx_{n} + \Delta \xx_{n}}}.
\]
We first consider the case that
  $\zz \not \in \ray{\cc} - \hull{\xx_{1}, \ldots , \xx_{n} }$.
In this case, we will show that
  $\dist{\zz}{\bdry{\ray{\cc} - \hull{\vp{\xx}{1}{n}}}} \leq \alpha$,
  contradicting assumption~(\ref{foo:3}).
Since $\zz  + \Delta \zz  \in
  \bdry{\ray{\cc + \Delta \cc } -
  \hull{\xx_{1} + \Delta \xx_{1}, \ldots , \xx_{n} + \Delta \xx_{n}}}$,
\[
\zz  + \Delta \zz  =
  \lambda (\cc + \Delta \cc ) -
  \sum_{i} \gamma_{i} (\xx_{i} + \Delta \xx_{i}),
\]
for some $\lambda \geq 0$ and
  $\vp{\gamma }{1}{n} \geq 0, \sum_{i} \gamma_{i} = 1$. We establish an
  upper bound on $\lambda$ by noting that
\begin{equation}\label{eqn:secondGeom1}
  \lambda
=
  \frac{
   \norm{ \zz  + \Delta \zz  +
     \sum_{i} \gamma_{i} (\xx_{i} + \Delta \xx_{i})}
  }{
   \norm{ \cc + \Delta \cc }
  }.
\end{equation}
We lower bound the denominator of \eqref{eqn:secondGeom1} by $\norm{\cc}/2$ by
\[
  \norm{\Delta \cc} \leq
   \frac{\alpha \norm{\cc}}
        {2 \alpha + 4 (\norm{\zz} + \max_{i} \norm{\xx_{i}})}
  \leq
    \norm{\cc}/2.
\]
We upper bound the numerator of \eqref{eqn:secondGeom1} by
\begin{eqnarray*}
  \norm{  \zz  + \Delta \zz  +
     \sum_{i} \gamma_{i} (\xx_{i} + \Delta \xx_{i})}
& \leq &
  \norm{\zz} + \alpha /4 +
       \sum_{i} \gamma_{i} (\norm{\xx_{i}} + \norm{\Delta \xx_{i}})\\
& \leq &
  \norm{\zz} + \alpha /4 +
       \max_{i}\norm{\xx_{i}} + \alpha / 4\\
& = &
  \norm{\zz} +
       \max_{i}\norm{\xx_{i}} + \alpha / 2.\\
\end{eqnarray*}
Thus,
\[
  \lambda
\leq
  \frac{
   \norm{\zz} +  \max_{i}\norm{\xx_{i}} + \alpha / 2
}{
   \norm{\cc}/2
}
\]
Since
\[
\left(\zz  + \Delta \zz  - \lambda \Delta \cc + \sum_{i} \gamma_i \Delta \xx_i \right)
  = \left( \lambda \cc - \sum_{i} \gamma_i \xx_i \right)
  \in \ray{\cc} - \hull{\xx_{1}, \ldots, \xx_{n}},
\]
we find that
\begin{eqnarray*}
\lefteqn{\dist{\zz}{\bdry{\ray{\cc} - \hull{\vp{\xx}{1}{n}}}}}\\
&  \leq  &
 \norm{ \Delta \zz - \lambda \Delta \cc + \sum_{i}\gamma_{i}
  \Delta\xx_{i}}
\\
 & \leq &
 \norm{\Delta \zz} + \lambda \norm{\Delta \cc} +
   \sum_{i}\gamma_{i} \norm{ \Delta \xx_{i}}\\
 & \leq &
  \frac{\alpha}{4}+
  \left(  \frac{
      \norm{\zz} +  \max_{i}\norm{\xx_{i}} + \alpha / 2
   }{
      \norm{\cc}/2
   }
  \right)
  \left(
   \frac{\alpha \norm{\cc}}
        {2 \alpha + 4 (\norm{\zz} + \max_{i} \norm{\xx_{i}})}
  \right)
   + \frac{\alpha}{4} \\
 & = & \alpha ,
\end{eqnarray*}
contradicting assumption~(\ref{foo:3}).

We now consider the case that
  $\zz \in \ray{\cc  } - \hull{\xx_{1}, \ldots , \xx_{n} }$.
Since
\[
  \zz  + \Delta \zz  \in
  \bdry{\ray{\cc + \Delta \cc } -
  \hull{\xx_{1} + \Delta \xx_{1}, \ldots, \xx_{n} + \Delta \xx_{n}}},
\]
  there exists a hyperplane $H$ passing through $\zz  + \Delta \zz$ and
  tangent to the convex set\\ $\ray{\cc + \Delta \cc } -
  \hull{\xx_{1} + \Delta \xx_{1}, \ldots, \xx_{n} + \Delta
  \xx_{n}}$. By the assumption that\\
  $\dist{\zz}{\bdry{\ray{\cc} - \hull{\vp{\xx}{1}{n}}}} > \alpha$,
  there is some $\delta_0 > 0$ such that, for every $\delta \in
  (0,\delta_0)$, every point within
  $\alpha+\delta$ of $\zz$ lies within
  $\ray{\cc} - \hull{\vp{\xx}{1}{n}}$. Choose $\delta \in (0,\delta_0)$
  that also satisfies $\delta \leq \norm{\zz} + \max_i \norm{\xx_i}$.
  Let $\qq$ be a point at distance
  $\frac{3\alpha}{4}+\delta$ from $\zz  + \Delta \zz$ in the direction
  perpendicular to $H$. Since $\dist{\zz}{\zz + \Delta \zz} \leq
  \frac{\alpha}{4}$, and $\dist{\zz + \Delta \zz}{\qq} =
  \frac{3\alpha}{4} + \delta$,
\[
  \qq \in \ray{\cc} - \hull{\xx_{1}, \ldots, \xx_{n}}
\]
At the same time,
\begin{equation}\label{eqn:secondGeom2}
  \dist{\qq}{\ray{\cc + \Delta \cc } -
  \hull{\xx_{1} + \Delta \xx_{1}, \ldots, \xx_{n} + \Delta \xx_{n}}}
  > \frac{3\alpha}{4}.
\end{equation}
Because $\qq \in \ray{\cc} - \hull{\xx_{1}, \ldots, \xx_{n}}$, there
  exist $\lambda \geq 0$ and
  $\vp{\gamma }{1}{n} \geq 0, \sum_{i} \gamma_{i} = 1$ such that
\[
  \qq = \lambda \cc - \sum_{i} \gamma_{i} \xx_{i}.
\]
We upper bound $\lambda$ as before,
\[
\lambda =
  \frac{
   \norm{ \qq + \sum_{i} \gamma_{i} \xx_{i}}
  }{
   \norm{ \cc}
  }
\leq
  \frac{
   \norm{ \zz} + \alpha +\delta + \max_{i} \norm{\xx_{i}}
  }{
   \norm{ \cc}
  }
\leq
  \frac{
   \norm{\zz} +  \max_{i}\norm{\xx_{i}} + \alpha / 2
}{
   \norm{\cc}/2
}
\]
Hence,
\[
  \qq + \lambda \Delta \cc - \sum_{i} \gamma_{i} \Delta \xx_{i} =
  \lambda (\cc + \Delta \cc) - \sum_{i} \gamma_{i} (\xx_{i} + \Delta
  \xx_{i})
\]
\[ \in~~
  \ray{\cc + \Delta \cc } -
  \hull{\xx_{1} + \Delta \xx_{1}, \ldots, \xx_{n} + \Delta \xx_{n}},
\]
and thus
\begin{eqnarray*}
\dist{\qq}
  {\ray{\cc + \Delta \cc } -
  \hull{\xx_{1} + \Delta \xx_{1}, \ldots, \xx_{n} + \Delta \xx_{n}}}
& \leq &
  \norm{\lambda \Delta \cc - \sum_{i} \gamma_{i} \Delta \xx_{i}}\\
& \leq &
  \lambda \norm{\Delta \cc} + \max_{i} \norm{\Delta \xx_{i}}\\
& \leq &
  \alpha/2 + \alpha/4\\
& \leq &
  3\alpha/4,
\end{eqnarray*}
which contradicts~\eqref{eqn:secondGeom2}.
\end{proof}

\subsection{Dual condition number is likely low}\label{ssec:dual-prob}

\begin{proof}[Proof of Lemma~\ref{lem:dual2}]
Let
\[
  \zz = \frac{1}{n}\sum _{i}\aa _{i}
  \mbox{ and }
  \xx _{i} = \aa _{i} - \zz ,
  \mbox{ for $1 \leq i \leq n$},
\]
\[
 k_1 = \dist{\zz}{\bdry{\ray{\cc} - \hull{\vp{\xx}{1}{n}}}} \mbox{ and }
 k_2 = \norm{\cc }.
\]
We will apply the bound of Lemma~\ref{lem:dualGeom}.
We first lower bound $\min\setof{k_1, k_2, k_1 k_2}$.
We begin by observing that if
\[
  \min\setof{k_1, k_2, k_1 k_2} < \epsilon,
\]
then either
\begin{eqnarray}
  \dist{\zz}{\bdry{\ray{\cc} -\hull{\vp{\xx}{1}{n}}}} < \epsilon,\label{eqn:foo1}
\end{eqnarray}
or
\begin{eqnarray}
  \norm{\cc} < \epsilon,\label{eqn:foo2}
\end{eqnarray}
or there exists some integer $l$, $1 \leq l \leq \ceiling{\log
\frac{1}{\epsilon}}$, for which
\begin{eqnarray}
  \dist{\zz}{\bdry{\ray{\cc} -\hull{\vp{\xx}{1}{n}}}} < 2^{l} \epsilon
  ~ \mbox{ and } ~
  \norm{\cc } \leq 2^{-l+1}. \label{eqn:foo3}
\end{eqnarray}
The probabilities of events \eqref{eqn:foo1} and \eqref{eqn:foo2} will also
be bounded in our
analysis of event \eqref{eqn:foo3}.
By Proposition~\ref{pro:chiSquare2}, for $d \geq 2$, we have
\[
  \prob{}{  \norm{\cc } \leq \epsilon}
  \leq \frac{e \epsilon}{\sigma},
\]
which translates to
\[
  \prob{}{  \norm{\cc } \leq 2^{-l+1}}
  \leq \frac{e 2^{-l+1}}{\sigma},
\]
while Lemma~\ref{lem:dualProb} implies
\[
   \prob{\zz}{  \dist{\zz}{\bdry{\ray{\cc} -\hull{\vp{\xx}{1}{n}}}} <2^{l} \epsilon }
     \leq \frac{8 \cdot 2^{l} \epsilon d^{1/4}n^{1/2}}{ \sigma} .
\]

Thus, we compute
\begin{eqnarray*}
 \prob{}
      {
    \min\setof{k_1, k_2, k_1 k_2}
    < \epsilon }
& \leq &
     \frac{8~  \epsilon d^{1/4}n^{1/2}}{ \sigma}
  + \frac{e \epsilon}{\sigma} + \sum_{l = 1}^{\ceiling{\log \frac{1}{\epsilon}}}
     \frac{e 2^{-l+1}}{\sigma}
     \frac{8 \cdot 2^{l} \epsilon d^{1/4}n^{1/2}}{ \sigma}\\
& = &
\frac{8~\epsilon d^{1/4}n^{1/2}}{\sigma} +
\frac{e \epsilon}{\sigma} +
\frac{16 e \epsilon d^{1/4}n^{1/2}}{\sigma^2}\ceiling{\log\left(\frac{1}{\epsilon}\right)}\\
& \leq &
  \frac{55~\epsilon d^{1/4}n^{1/2}}{\sigma^2}\ceiling{\log\left(\frac{1}{\epsilon}\right)} .
\end{eqnarray*}
Setting 
\[
   \epsilon = \frac{\delta}
             {200  d^{1/4} n^{1/2}
            \log \left(\frac{200 d^{1/4}
               n^{1/2}}{\sigma^{2} \delta } \right) \big/ \sigma^{2}},
\]
we find that 
\[
  \frac{55~\epsilon d^{1/4}n^{1/2}}{\sigma^2}\log\left(\frac{1}{\epsilon}\right) 
\leq \delta /2,
\]
for $\delta \leq 1$.  So, we obtain
\[
 \prob{}
      {
    \min\setof{k_1, k_2, k_1 k_2}
    < 
 \frac{\delta}
             {200   d^{1/4} n^{1/2}
          \log \left(\frac{200 d^{1/4}
               n^{1/2}}{\sigma^{2} \delta } \right) \big/ \sigma^{2}}
 }
  < \delta /2,
\]
which we re-write this as
\begin{eqnarray}\label{eqn:dual1}
 \prob{}
      { \max \setof{\frac{1}{k_{1}}, \frac{1}{k_{2}}, \frac{1}{k_{1}k_{2}}}
    >
    \frac{200~d^{1/4} n^{1/2}}{\delta  \sigma^2}
    \log\left(\frac{200~d^{1/4} n^{1/2}}{\delta \sigma^2}\right)
    }   < \frac{\delta}{2}.
\end{eqnarray}

By Lemma \ref{lem:dualGeom}, we have
\begin{eqnarray*}
  \frac{1}{\rho(A, \cc)} & \leq &
\max\setof{\frac{8}{k_1}, \frac{4}{k_2}, \frac{24
\max_i{\norm{\aa_i}}}{k_1 k_2}}\\
& \leq & 24\max (\max_{i}\norm{\aa_{i}}, 1)
\max \setof{\frac{1}{k_{1}}, \frac{1}{k_{2}}, \frac{1}{k_{1}k_{2}}}.
\end{eqnarray*}

By Corollary~\ref{cor:chiSquare}, we have

\begin{eqnarray}\label{eqn:dual2}
  \prob{}{\max(\fnorm{A, \cc}, 1)
            > 3 + \sigma \sqrt{(d+1) n~ 2 \ln (4e/\delta)}}
   < \frac{\delta}{4},
\end{eqnarray}
and $\max(\fnorm{A, \cc}, 1) \geq \max (\max_{i}\norm{\aa_{i}}, 1)$.

From a union bound of inequalities \ref{eqn:dual1} and
  \ref{eqn:dual2}, we obtain

%Thus, by a union bound using inequalities \ref{eqn:dual1},
%  \ref{eqn:dual2}, and \ref{eqn:dual3}, we obtain

\begin{eqnarray*}
\prob{}{
      \frac{\fnorm{A, \cc}}{\rho(A, \cc)} >
      \frac{24\cdot200~d^{1/4} n^{1/2}}{\delta  \sigma^2}
    \log\left(\frac{200~d^{1/4} n^{1/2}}{\delta \sigma^2}\right)
 (3 + \sigma \sqrt{(d+1) n 2 \ln (2e/\delta)})^{2}
}
& \leq & \delta.
\end{eqnarray*}
The proof of the first part of the lemma now follows by computing
$$\frac{24\cdot200~d^{1/4} n^{1/2}}{\delta  \sigma^2}
    \log\left(\frac{200~d^{1/4} n^{1/2}}{\delta \sigma^2}\right)
  (3 + \sigma \sqrt{(d+1) n~ 2 \ln (2e/\delta)})^{2} \leq$$
$$  \frac{50000~ d^{1/4} n^{1/2}}{\delta \sigma^2}
  \log^{2} \left( \frac{200~d^{1/4} n^{1/2}}{\delta \sigma^2}
\right),$$
where we used the assumption $\sigma \leq 1/\sqrt{dn}$.

We now establish the smoothed bound on the log of expectation.
Note that
\begin{eqnarray*}
\expec{}
   {\log \frac{\fnorm{A, \cc}}{\rho(A, \cc )}}
& = & \expec{}{\log \fnorm{A, \cc}} +
      \expec{}{\log \frac{1}{\rho(A, \cc )}} \\
&\leq &
  \expec{}{\log \fnorm{A, \cc}} +
  \expec{}{\log \max \setof{\frac{1}{k_{1}}, \frac{1}{k_{2}},
   \frac{1}{k_{1}k_{2}}}}  +
  \expec{}{\log (24\max (\max_{i}\norm{\aa_{i}},1))}\\
& \leq & \expec{}{\log \max (\fnorm{A, \cc},1)} +
  \expec{}{\log \max \setof{\frac{1}{k_{1}}, \frac{1}{k_{2}},
   \frac{1}{k_{1}k_{2}}}} + \expec{}{24\log \max (\fnorm{A},1)} \\
& \leq  & \log{\sqrt{n (d+1)+1}} + 2 \log
    \frac{55d^{1/4}n^{1/2}}{\sigma^{2}} + 2 + \log 24 + \log{\sqrt{nd+1}}
\\
& \leq  & 14 + 4 \log \frac{nd}{\sigma },
\end{eqnarray*}
where the bound is derived using the same argument as in the
 proof of Lemma~\ref{lem:logcond}.

\end{proof}

% Local Variables: ***
% TeX-master:"journalipm.tex" ***
% End: ***

\section{Combining the Primal and Dual Analyses}\label{sec:cond}

\begin{proof}[Proof-of-theorem~\ref{thm:main}]

Note that the transformation of each canonical form into the conic form
  leaves the Frobenius norm unchanged.
Also, a random Gaussian perturbation in the original form maps to
  a random Gaussian perturbation in the conic form.
Therefore, by Lemma~\ref{lem:homogenizing}, 
  the smoothed bounds on the primal and dual condition numbers
  of the conic forms imply  smoothed bounds
  on each of the condition
  numbers $C_{P}^{(1)}$, $C_{P}^{(2)}$,
  $C_{D}^{(2)}$, $C_{D}^{(3)}$.

By Lemmas \ref{lem:primal} and Lemma~\ref{lem:dual2}, we have
  that for all $\orig{A}$, $\orig{\bb}$ and $\orig{\cc}$
  satisfying $\fnorm{\orig{A}, \orig{\bb}, \orig{\cc }} \leq 1$
  and $\sigma \leq 1/\sqrt{nd}$,
\begin{align*}
\lefteqn{ 
  \prob{A,\bb ,\cc }
       {C^{(i)} (A,\bb ,\cc )
            >
    \frac{2^{13}~ (n+1)^{2} (d+1)^{1.5} }{\delta \sigma^2}\left(\log
    \frac{2^{10}~(n+1)^2 (d+1)^{1.5}}{\delta \sigma^2} \right)^2
}
}  \\
& \leq  
  \prob{A,\bb ,\cc }
       {C_{P}^{(i)} (A,\bb)
            >
    \frac{2^{12}~ n^{2} (d+1)^{1.5} }{(\delta/2) \sigma^2}\left(\log
    \frac{2^{9}~n^2 (d+1)^{1.5}}{(\delta/2) \sigma^2} \right)^{2}
}
\\ 
& \qquad +
  \prob{A,\bb ,\cc }
       {C_{D}^{(i)} (A,\cc)
            >
    \frac{2^{12}~ (n+1)^{2}d^{1.5} }{(\delta/2) \sigma^2}\left(\log
    \frac{2^{9}~(n+1)^2d^{1.5}}{(\delta/2) \sigma^2} \right)^{2}
}\\
&
\leq  \delta/2 + \delta /2 = \delta.
\end{align*}

To bound the log of the condition number, we use
  Lemmas~\ref{lem:logcond} and Lemma~\ref{lem:dual2} to show
\begin{align*}
\lefteqn{
  \expec{A,\bb ,\cc }
       {\log C^{(i)}(A,\bb ,\cc )}
}
 \\
  & \leq 
  \expec{A,\bb ,\cc }
       {\log \left (C^{(i)}_{P} (A,\bb) + C^{(i)}_{D} (A,\cc ) \right)
}
   \\
& \leq 
 \max\left (
  \expec{A,\bb ,\cc }
       {\log \left (2C^{(i)}_{P} (A,\bb)
    \right)}, \expec{A,\bb ,\cc }
       {\log \left (2 C^{(i)}_{D} (A,\cc ) \right)}
  \right) \\
& \leq  15 + 4.5 \log \left(\frac{nd}{\sigma }\right),
\end{align*}
where in the second-to-last inequality used that fact that  for positive
  random variables $\beta $ and $\gamma $,
\[
\expec{}{\log (\beta +\gamma  )} \leq \max \left(\expec{}{\log (2\beta )},\expec{}{\log (2\gamma )}\right).
\]
\end{proof}

% Local Variables: ***
% TeX-master:"journalipm.tex" ***
% End: ***

%\input{renegar-chapter-5-and-6}
\section{Open Problems and Conclusion}\label{sec:discussion}

The best way to strengthen the results in this paper would
  be to prove that they hold under more restrictive
  models of perturbation.
For example, we ask whether similar results can be proved
  if one perturbs the linear program subject to maintaining
  feasibility or infeasibility.
This would be an example of a property-preserving perturbation,
  as defined in~\cite{SpielmanTengProperty}.

A related  question is whether these results can be proved under
  zero-preserving perturbations in which only non-zero entries of $A$
  are subject to perturbations.
Unfortunately, the following example shows that in this model
  of zero-preserving perturbations,
  it is not possible to bound the condition number by
  $poly(n,d,\frac{1}{\sigma})$ with probability at least $1/2$.
Therefore, if such a result were to hold in the model of zero-preserving perturbations,
  it would not be because
  of a polynomial bound on the condition number.

Let $A$ be a zero preserving Gaussian perturbation of
  $\orig{A}$ with variance $\sigma^2$.
For ease of exposition, we will normalize $\fnorm{\orig{A}}$ to
  be 1 at the end of formulation.
Define the matrix
$$\orig{A} = \begin{bmatrix}
-1 & \epsilon & & \\
  & -1 & \epsilon &\\
& & \cdots\\
& & -1 & \epsilon
\end{bmatrix}$$
  where $\epsilon$ is a parameter to be chosen later, and consider the linear
  program $A\xx \geq \bvec{0}, \xx \in \CC$
  where $\CC = \{\xx: \xx > \bvec{0} \}$.
The $i^{th}$ constraint of $\orig{A} \xx \geq \bvec{0}$ is
  exactly $$\epsilon x_{i+1} \geq  x_i$$
We apply fact~\ref{fact:2} with $c=\delta^2/\sigma^2$
  assumed to be at least 6 (so that $(1-c+\ln c) \leq -c/2$).
This yields
\begin{eqnarray}
\Pr[|a_{i,i}-1| \geq \delta] \leq e^{-\frac{1}{2} (1 -
\frac{{\delta}^2}{\sigma^2} + \ln \frac{{\delta}^2}{\sigma^2})}
\leq e^{-\frac{{\delta}^2}{4 \sigma^2}}
\label{eqn:bar1}\\
\Pr[|a_{i,i+1} - \epsilon| \geq \delta] \leq e^{-\frac{1}{2} (1 -
\frac{{\delta}^2}{\sigma^2} + \ln \frac{{\delta}^2}{\sigma^2})}
\leq e^{-\frac{{\delta}^2}{4 \sigma^2}} \label{eqn:bar2}
\end{eqnarray}
Setting $\delta = \sigma \sqrt{8\log n}$ yields that,
  with probability at least $1/2$,
  none of the events~(\ref{eqn:bar1}),~(\ref{eqn:bar2}) happen for any $i$.
Assuming that none of the
events~(\ref{eqn:bar1}),~(\ref{eqn:bar2}) occur, and that
$\epsilon > \delta$ (which we will ensure later),
  we have that $A\xx \geq
  \bvec{0}, \xx \in \CC$ is feasible, and
  $$\xx = \left[\left(\frac{\epsilon-\delta}{1+\delta}\right)^{n},
\left(\frac{\epsilon-\delta}{1+\delta}\right)^{n-1}, ~\ldots ~,
1\right]$$ is one such feasible solution. We also have that
$(\epsilon + \delta) x_{i+1} \geq (1-\delta) x_i$ for every $i$.
Define
$$\Delta A = \begin{bmatrix}
0 & \ldots & 0 &  -(\frac{\epsilon+\delta}{1-\delta})^{n-2}\\
0 & \ldots & 0 & 0\\
& \cdots\\
0 & \ldots & 0 & 0
\end{bmatrix}.
$$
We now show that $(A+\Delta A)\xx \geq \bvec{0}, \xx \in \CC$ is
infeasible, and hence $\rho(A,\CC) \leq \fnorm{\Delta A} =
(\frac{\epsilon+\delta}{1-\delta})^{n-2}$. To see infeasibility,
note that the constraint given by the top row of $(A+\Delta A)$ is
$$-x_1 + \epsilon x_2 - \left(\frac{\epsilon+\delta}{1-\delta}\right)^{n-2}x_n \geq 0$$
while we simultaneously have that $x_2 \leq
(\frac{\epsilon+\delta}{1-\delta})^{n-2} x_n$. Assuming $\epsilon
\leq 1$ (which we ensure later), this constraint is impossible to
satisfy for $\xx \in \CC$.

Letting $\epsilon = \frac{1}{n}$ and $\sigma = \frac{1}{n^2}$
  (and hence $\delta = \frac{\sqrt{8 \log n}}{n^2}$)
  yields $\rho(A,\CC) = (\frac{\epsilon+\delta}{1-\delta})^{n-2}
  = (\frac{O(1)}{n})^{n-2}$, which is exponentially small
  and also satisfies the requirements on $\epsilon$. We can upper bound $\fnorm{A}$
  by $\fnorm{A} \leq \sqrt{n(1+\delta)^2 + n(\epsilon+\delta)^2} \leq 2\sqrt{n}$.
Thus the condition number, which is equal to
$\fnorm{A}/\rho(A,\CC)$, is at
 least $\Omega(n)^{n-3}$.

If we had normalized $\fnorm{\orig{A}} = 1$ at the beginning of
the proof,
  the corresponding normalization would have been
  $\epsilon \approx \frac{1}{n\sqrt{n}}$, $\sigma \approx
  \frac{1}{n^2 \sqrt{n}}$, which still shows the negative result.
This analysis also shows the impossibility of a theorem like
  theorem~\ref{thm:main} for another natural model of perturbation,
  {\em relative perturbation}, that is also zero-preserving:
   multiplying each entry of
  $\orig{A}$ by an $N(1,\sigma^2)$
  Gaussian random variable.
This concludes our discussion of impossibility results for smoothed analysis.

We would like to point out that condition numbers appear
  throughout Numerical Analysis and that
  condition numbers may be defined for many non-linear problems.
The speed of algorithms for optimizing linear functions over
convex bodies
  (including semidefinite programming) has been related to
  their condition numbers~\cite{FreundConvex,FreundVera},
  and it seems that one might be able to extend
  our results to these algorithms as well.
Condition numbers have also been defined for non-linear
programming
  problems, and one could attempt to perform a smoothed analysis of
  non-linear optimization algorithms by relating their performance
  to the condition numbers of their inputs, and then performing
  a smoothed analysis of their condition numbers.

The approach of proving smoothed complexity bounds by relating
  the performance of an algorithm to some property of its input,
  such as a condition number, and the performing a smoothed
  analysis of this quantity has also been recently
  used in~\cite{SpielmanTengProperty,SankarSpielmanTeng}.
Finally, we hope that this work illuminates some of the shared
  interests of the Numerical Analysis, Operations Research,
  and Theoretical Computer Science communities.

\bibliographystyle{alpha}
\bibliography{journalipm}
\appendix
\section{Gaussian random variables}\label{sec:gaussian}

We now derive particular versions of well-known 
  bounds on the Chi-Squared distribution. 
These bounds are used in the body of the paper, and bounds of this form are
well-known. We thank DasGupta and Gupta~\cite{DasguptaGupta} 
for this particular derivation.

\begin{fact}[Sum of gaussians]\label{fact:sum}
Let $X_1, \ldots, X_d$ be independent $N(0,\sigma)$ random variables. Then 
$$\Pr[\sum_{i=1}^d X_i^2 \geq \kappa^2] \leq 
e^{ \frac{d}{2}(1 - \frac{\kappa^2}{d\sigma^2} + \ln
\frac{\kappa^2}{d\sigma^2})}$$ 
\end{fact}

\begin{proof} For simplicity, we begin with $Y_i
\sim N(0,1)$. A simple integration shows that if $Y \sim N(0,1)$ then
$E[e^{tY^2}] = \frac{1}{\sqrt{1-2t}} \quad (t < \frac{1}{2})$. We proceed with

\begin{eqnarray*}
\Pr[\sum_{i=1}^d Y_i^2 \geq k] & = &\\
\Pr[\sum_{i=1}^d Y_i^2 - k \geq 0] & = & \quad (\text{for } t >0)\\
\Pr[e^{t(\sum_{i=1}^d Y_i^2 -k)} \geq 1] & \leq & \quad (\text{by Markov's Ineq.})\\
\expec{}{e^{t(\sum_{i=1}^d Y_i^2 -k)}} & = &\\ 
\left(\frac{1}{1-2t}\right)^{d/2} e^{-kt} & \leq & \quad
(\text{letting } t = \frac{1}{2} - \frac{d}{2k})\\
\left(\frac{k}{d}\right)^{d/2} e^{-\frac{k}{2}+\frac{d}{2}} & = &
e^{\frac{d}{2}(1 - \frac{k}{d} + \ln \frac{k}{d} )}
\end{eqnarray*}

Since 
$$\Pr[\sum_{i=1}^d Y_i^2 \geq k] = 
\Pr[\sum_{i=1}^d X_i^2 \geq \sigma^2 k]$$ 
we set $k = \frac{\kappa^2}{\sigma^2}$ and obtain
$e^{\frac{d}{2}(1 - \frac{k}{d} + \ln \frac{k}{d} )} =
e^{ \frac{d}{2}(1 - \frac{\kappa^2}{d\sigma^2} + \ln
\frac{\kappa^2}{d\sigma^2})}$
which was our desired bound.
\end{proof}

In particular, this implies:
\begin{fact}[Alternative sum of gaussians]\label{fact:2}
Let $X_1, \ldots, X_d$ be independent $N(0,\sigma)$ random variables. Then 
for $c \geq 1$,
\[
\Pr[\sum_{i=1}^d X_i^2 \geq cd\sigma^2] \leq 
e^{\frac{d}{2}(1-c+\ln c)}.
\]
\end{fact}

\begin{corollary}\label{cor:chiSquare}
Let $\xx$ be a $d$-dimensional Gaussian random vector of
  variance $\sigma^{2}$ centered at the origin.
Then, for $d \geq 2$ and $\epsilon \leq 1/e^2$,
\[
  \prob{}{\norm{\xx} \geq \sigma \sqrt{d (1+2\ln (1 / \epsilon) }} \leq
  \epsilon
\]
\end{corollary}
\begin{proof}
Set $c = 1+2\ln (1 / \epsilon)$ in fact~\ref{fact:2}.
We then compute
$$e^{\frac{d}{2}(1-c+\ln c)} \leq e^{1-c+\ln c} \leq 
e^{-2 \ln \frac{1}{\epsilon} + \ln (1+2\ln \frac{1}{\epsilon})}
= \epsilon e^{- \ln \frac{1}{\epsilon} + \ln (1+2\ln \frac{1}{\epsilon})}$$
We now seek to show
\begin{eqnarray*}
e^{- \ln \frac{1}{\epsilon} + \ln (1+2\ln \frac{1}{\epsilon})} 
& \leq & 1\\
\Leftrightarrow \quad
- \ln \frac{1}{\epsilon} + \ln (1+2\ln \frac{1}{\epsilon})
& \leq & 0\\
\Leftrightarrow \quad
1+2\ln \frac{1}{\epsilon} & \leq & \frac{1}{\epsilon}
\end{eqnarray*}
For $\epsilon = 1/e^2$, the left-hand side of the last inequality is 5,
while the right-hand side is greater than 7. Taking derivatives with respect to
$1/\epsilon$, we see that the right-hand side grows faster as we
increase $1/\epsilon$ (decrease $\epsilon$), 
and therefore will always be greater.
\end{proof}

We also use the following easy-to-prove fact,
  a proof of which may be found in~\cite[Proposition~2.4.7]{SpielmanTeng}
\begin{proposition}\label{pro:chiSquare2}
Let $\xx$ be a $d$-dimensional Gaussian random vector of
  variance $\sigma^{2}$ centered at the origin. Then, 
\[
  \prob{}{\norm{\xx} \leq \epsilon} \leq
  \left(\frac{ \epsilon}{\sigma} \right)^{d}.
\]
\end{proposition}

% Local Variables: ***
% TeX-master:"journalipm.tex" ***
% End: ***

\end{document}